\documentclass{aa}

\usepackage{graphicx}
\usepackage{natbib}
\usepackage{mathrsfs}
\bibpunct{(}{)}{;}{a}{}{,}
 
\begin{document}

\title{On the internal structure of starless cores}
\subtitle{II. A molecular survey of L1498 and L1517B}

\author{M. Tafalla \inst{1} 
\and J. Santiago \inst{1}
\and P.C. Myers \inst{2}
\and P.Caselli \inst{2,3}
\and C.M. Walmsley \inst{3}
\and A. Crapsi \inst{4}
}
 
\institute{Observatorio Astron\'omico Nacional, Alfonso XII 3, E-28014 Madrid,
Spain
\and
Harvard-Smithsonian Center for Astrophysics, 60 Garden St., Cambridge,
MA 02138, USA
\and
Osservatorio Astrofisico di Arcetri, Largo E. Fermi 5, I-50125
Firenze, Italy
\and
Leiden Observatory, P.O. Box 9513, 2300 RA Leiden, The Netherlands
}
 
\offprints{M. Tafalla \email{m.tafalla@oan.es}}
\date{Received -- / Accepted -- }
 
\abstract
%Context
{Low mass starless cores present an inhomogeneous chemical composition. 
Species like CO and CS deplete at their dense interiors, while
N$_2$H$^+$ and NH$_3$ survive in the gas phase. As molecular line
observations are used to determine the physical conditions and kinematics 
of the core gas, chemical inhomogeneities can introduce a serious bias.}
%Aims
{We have carried out a molecular survey towards two
starless cores, L1498 and L1517B.
These cores have been selected for their relative isolation and
close-to-round shape. They have been observed in a number of lines of
13 molecular species in order to determine a self-consistent set of
abundance profiles.}
%Methods
{In a previous paper we modeled the physical structure
of L1498 and L1517B. Here we use this work
together with a spherically-symmetric Monte Carlo
radiative transfer code to determine the radial profile
of abundance for each species in the survey. Our model
aims to fit simultaneously the radial profile of integrated
intensity and the emerging spectrum from the core center.}
%Results
{L1498 and L1517B present similar abundance patterns,
with most species suffering a significant drop
toward the core center. This occurs for 
CO, CS, CH$_3$OH, SO, C$_3$H$_2$, HC$_3$N, C$_2$S, HCN, H$_2$CO,
HCO$^+$, and DCO$^+$, which we fit with profiles
having a sharp central hole. The size of this hole varies with
molecule: DCO$^+$, HCN, and HC$_3$N have the smallest holes,
while SO, C$_2$S and CO have the largest holes. Only N$_2$H$^+$ 
and NH$_3$ seem present in the gas phase at the core centers.} 
%Conclusions.
{From the different behavior of molecules, we select
SO, C$_2$S, and CH$_3$OH as the most sensitive tracers of molecular
depletion. Comparing our abundance determinations with the predictions 
from current chemical models we find order of
magnitude discrepancies. Finally, we show how the
``contribution function'' can be used to study the formation 
of line profiles from the different regions of a core.}

\keywords{ISM: abundances -- ISM: clouds --ISM: molecules -- 
stars: formation -- ISM: individual(L1498, L1517B)}

\authorrunning{Tafalla et al.}
\titlerunning{On the internal structure of starless dense cores. II.}
 
\maketitle

\section{Introduction}

Recent observations and modelling of low-mass starless
cores show that a number of molecular species deplete
from the gas phase
at the dense interior of these simplest star-forming regions
\citep{kui96,wil98,kra99,alv99,cas99,ber01,taf02,bac02,pag05}. 
The depletion of molecules in the 
pre-stellar material has important consequences for the study of the 
initial conditions of stellar birth. Molecular emission is
routinely used to trace the physical properties and the kinematics
of the star-forming gas, so the removal of certain species from 
the gas phase introduces a potential distortion in molecular
line observations. Depletion, in addition, systematically
increases with time and gas density. Thus, if properly understood, it can
provide a reliable clock to time the contraction history of dense cores.
Understanding the chemical composition of starless cores
has therefore become necessary to reconstruct the earliest
phases of low-mass star formation.

Previous studies of core chemical composition show
that molecular depletion is a highly selective process. Molecules
like CO and CS disappear rapidly from the gas phase, while species like
N$_2$H$^+$ and NH$_3$ survive much longer at high densities (see 
references above). As a result of this behavior, a core gradually
develops a differentiated interior characterized by a center rich
in depletion-resistant species surrounded by layers richer in
depletion-sensitive molecules. Understanding this abundance pattern is 
critical to interpret molecular line observations
of cores because different species will systematically trace different 
layers depending both on their response to depletion and 
their level excitation.

Chemical models already provide an indication of how 
the different molecular species will be distributed inside a core,
but systematic molecular surveys are still needed to 
obtain a realistic picture of the core chemical composition.
Such surveys should be carried out toward starless cores of
simple geometry, so the abundances of the different species can
be unambiguously derived from observations. In this
paper, we present a survey toward two Taurus-Auriga
cores, L1498 and L1517B. These two cores are reasonably isolated and
present close to round shapes when observed in the millimeter
continuum, so they constitute ideal targets for a systematic study. 
In \citet{taf04a} (paper 1 hereafter), we used millimeter continuum 
data together with C$^{18}$O, CS, N$_2$H$^+$, and NH$_3$ 
line observations of these cores to derive their radial profiles of density, 
temperature, turbulent linewidth, and line-of-sight velocity. Now 
we complement this study with additional observations of a number of
molecules, and use the already derived physical models of the cores to 
determine in a self consistent manner the radial abundance distribution
of 13 molecular species. 

In the following
sections, we present the details of our radiative transfer modeling
of the observed lines and the set of abundance profiles
derived from this analysis (section 4). Using these abundances,
we study the differences and similarities between the two cores,
as well as their chemical relation with other cores of well
determined abundances (sections 5.1-5.3). We also use our data 
to test core chemical models, in particular
the recent work by \citet{aik05} (section 5.4), and to study the
different sensitivity of molecules to depletion (section 5.5).
We finish using our chemical determinations and the 
radiative transfer modeling to study how the different molecular lines 
originate at the core interior and therefore reflect
(or miss) its internal structure when used to trace the core gas
(section 5.6). For this analysis we use as a tool the {\em contribution
function.}

\section{Observations}

We observed L1498 and L1517B with the IRAM 30m telescope during several runs
between 1999 October and 2002 November. We made maps of these cores in the
lines shown in Table 1, always observing in frequency switching mode with
several receivers simultaneously and a typical sampling of $20''$. 
As a backend, we used a facility 
autocorrelator that provided a typical velocity resolution of 
0.03-0.04 km s$^{-1}$. During the observations, the telescope pointing
was corrected making frequent cross scans of bright continuum sources,
and the atmospheric attenuation was calibrated observing 
ambient and liquid nitrogen loads. The telescope intensity scale was
converted into main beam brightness temperature using standard
efficiencies. The FWHM of the telescope beam varied with frequency
from $27.7''$ at 90 GHz to $11''$ at 230 GHz.

We observed L1498 and L1517B in HCO$^+$(1--0), H$^{13}$CO$^+$(1--0),
and HCN(1--0) with the FCRAO 14m telescope\footnote
{FCRAO is supported in part by the National Science Foundation
under grant AST 94-20159, and is operated with permission of the Metropolitan
District Commission, Commonwealth of Massachusetts.}
in 2001 April. We used the QUARRY array receiver in frequency switching 
mode together with the facility correlator, which provided a
velocity resolution between 0.03 and 0.07 km s$^{-1}$. Observations
of SiO masers were used to correct the telescope pointing, and 
an ambient 
load was used to calibrate the atmospheric attenuation. An efficiency
of 0.55 was used to convert the telescope units into mean beam 
brightness temperature. The typical sampling of the maps was $30''$,
and the FWHM of the telescope beam at 86-89 GHz was approximately $55''$.

\begin{table}
\caption[]{Observed lines and rest frequencies used in this work.
\label{tbl-1}}
\[
\begin{array}{lcrrc}
\hline
\noalign{\smallskip}
\mbox{Line} & \mbox{Telescope} & \mbox{Frequency}  & E_l & \mbox{Ref.} \\
& & \mbox{(MHz)} & \mbox{(K)} &  \\
\noalign{\smallskip}
\hline
\noalign{\smallskip}
\mbox{CH$_3$OH(J$_k$=2$_0$--1$_0$)A$^+$~~~~~~} & \mbox{~~IRAM~~} & 
96741.375 & 2.3 & (1) \\
\mbox{CH$_3$OH(J$_k$=3$_0$--2$_0$)A$^+$} & \mbox{IRAM} & 145103.200 & 6.7 & (2) \\
\mbox{CH$_3$OH(J$_k$=2$_{-1}$--1$_{-1}$)E} & \mbox{IRAM} & 96739.362 & 0.0 & (1) \\
\mbox{CH$_3$OH(J$_k$=3$_{-1}$--2$_{-1}$)E} & \mbox{IRAM} & 145097.450 & 4.6 & (2) \\
\mbox{SO(JN=32--21)} & \mbox{IRAM} & 99299.890 & 4.5 & (2) \\
\mbox{SO(JN=43--32)} & \mbox{IRAM} & 138178.670 & 9.2 & (2) \\
\mbox{c-C$_3$H$_2$(J$_{K_{-1}K_{+1}}$=2$_{12}$--1$_{01}$)} & \mbox{IRAM} 
& 85338.894 & 0.0 & (3) \\
\mbox{c-C$_3$H$_2$(J$_{K_{-1}K_{+1}}$=3$_{12}$--2$_{21}$)} & \mbox{IRAM} 
& 145089.610 & 6.7 & (3) \\
\mbox{H$_2$CO(J$_{K_{-1}K_{+1}}$=2$_{12}$--1$_{11}$)} & \mbox{IRAM} 
& 140839.502 & 0.0 & (4) \\
\mbox{H$_2$CO(J$_{K_{-1}K_{+1}}$=2$_{11}$--1$_{10}$)} & \mbox{IRAM} 
& 150498.334 & 0.3 & (5) \\
\mbox{HC$_3$N(J,F=4,5--3,4)} & \mbox{100m} & 36392.363 & 2.6 & (4) \\
\mbox{HC$_3$N(J,F=10,11--9,10)} & \mbox{IRAM} & 90979.002 & 19.7 & (4) \\
\mbox{HCO$^+$(J=1--0)}  & \mbox{FCRAO} & 89188.523 & 0.0 & (5) \\
\mbox{HCO$^+$(J=3--2)}  & \mbox{IRAM} & 267557.619 & 12.9 & (5) \\
\mbox{H$^{13}$CO$^+$(J=1--0)}  & \mbox{FCRAO} & 86754.288 & 0.0 & (6) \\
\mbox{DCO$^+$(J=3--2)}  & \mbox{IRAM} & 216112.582 & 10.4 & (7) \\
\mbox{C$_2$S(JN=67--56)}  & \mbox{IRAM} & 86181.391 & 19.2 & (8) \\
\mbox{HCN(JF=12--01)}  & \mbox{FCRAO} & 88631.847 & 0.0 & (4) \\
\mbox{HCN(JF=34--23)}  & \mbox{IRAM} & 265886.487 & 12.8 & (4) \\
\mbox{H$^{13}$CN(JF=12--01)}  & \mbox{IRAM} & 86340.168 & 0.0 & (4) \\
\noalign{\smallskip}
\hline
\end{array}
\]
\begin{list}{}{}
\item[References:] (1) \citet{mul04}; (2) this work; (3) 
C. Gottlieb, priv. comm.; (4) CDMS; (5) JPL; (6) \citet{sch04};
(7) \citet{cas05}; (8) \citet{yam90}
\end{list}
\end{table}

For the detailed line modeling presented here, accurate rest frequencies
are required. We have searched for such frequencies using
on-line compilations like the Cologne
Database for Molecular Spectroscopy (CDMS, \citealt{mul01}) 
and the JPL Catalog \citep{pic98}. In most cases, the CDMS and 
JPL frequencies are consistent with each other, 
so we have chosen the one quoted as having smaller uncertainty (where 
possible, we have referred to the original determination). For
HCO$^+$, however, the values in the two catalogs are inconsistent,
probably due to the difficulty measuring frequencies 
in ions, and we have preferred the JPL value for its better 
agreement with our line data. For some transitions of CH$_3$OH
and SO, no accurate frequencies were found, and we have
determined them by fitting our L1498 spectra assuming
an LSR velocity of 7.80 km s$^{-1}$ (as measured from the other
lines). Fortunately, these few spectra present gaussian lines, so
the astronomical determination is likely to be accurate (about 20 kHz). A
summary of the
frequencies used in this work is presented in Table 1.

\section{Overview of the molecular emission}

In Figures 1 and 2 we present the integrated intensity maps of all
the lines observed toward L1498 and L1517B together with maps of the
dust continuum emission and N$_2$H$^+$(1--0) already analyzed in paper 1.
For each line, the integrated intensity
map reflects the combined effect of excitation, optical
depth, and molecular abundance, so its interpretation
requires detailed radiative transfer modeling. Even without
such an analysis we can 
appreciate the need for strong abundance 
variations by noting that most lines are
not very optically thick and that their excitation 
increases with density toward the dust continuum peak.
Thus, if the abundance of a molecule were spatially constant, its
emission map would present a well-defined maximum at the dust peak.
Although this occurs for N$_2$H$^+$ (whose abundance is close to 
constant, paper 1), it is not the case for the rest of the species. 

In the larger and better resolved
L1498 core, the maps of C$_3$H$_2$, H$_2$CO, CH$_3$OH, SO,
HCO$^+$ (plus isotopes), HCN (plus isotope), HC$_3$N, and
C$_2$S all present distributions that differ from the centrally 
concentrated dust or N$_2$H$^+$. In a few species, like CH$_3$OH, SO, 
and H$_2$CO, the emission forms an almost-complete ring around the 
dust peak. In others, the emission presents discrete peaks,
especially to SW and a NE of the dust peak, reminiscent of a broken 
ring. Such ring-like distributions are also seen
in C$^{18}$O and CS, and in paper 1 it was shown 
that they reflect the presence of a central depletion hole 
(see also \citealt{kui96, wil98}). 
The maps in Fig. 1 show now that depletion holes
must be the rule for most species in L1498.

For the more compact L1517B core, the pattern of line emission
is similar to that in L1498, although less distinct for some
species because of angular resolution. 
In this core, most molecules present a single emission peak 
offset to the west from the dust/N$_2$H$^+$ peak, being the most
extreme example that of SO. This pattern is again similar to that found
in C$^{18}$O and CS (paper 1), and for the same reasons as in L1498,
it requires a central depletion hole.

\begin{figure*}[t]
\centering
\resizebox{15cm}{!}{\includegraphics{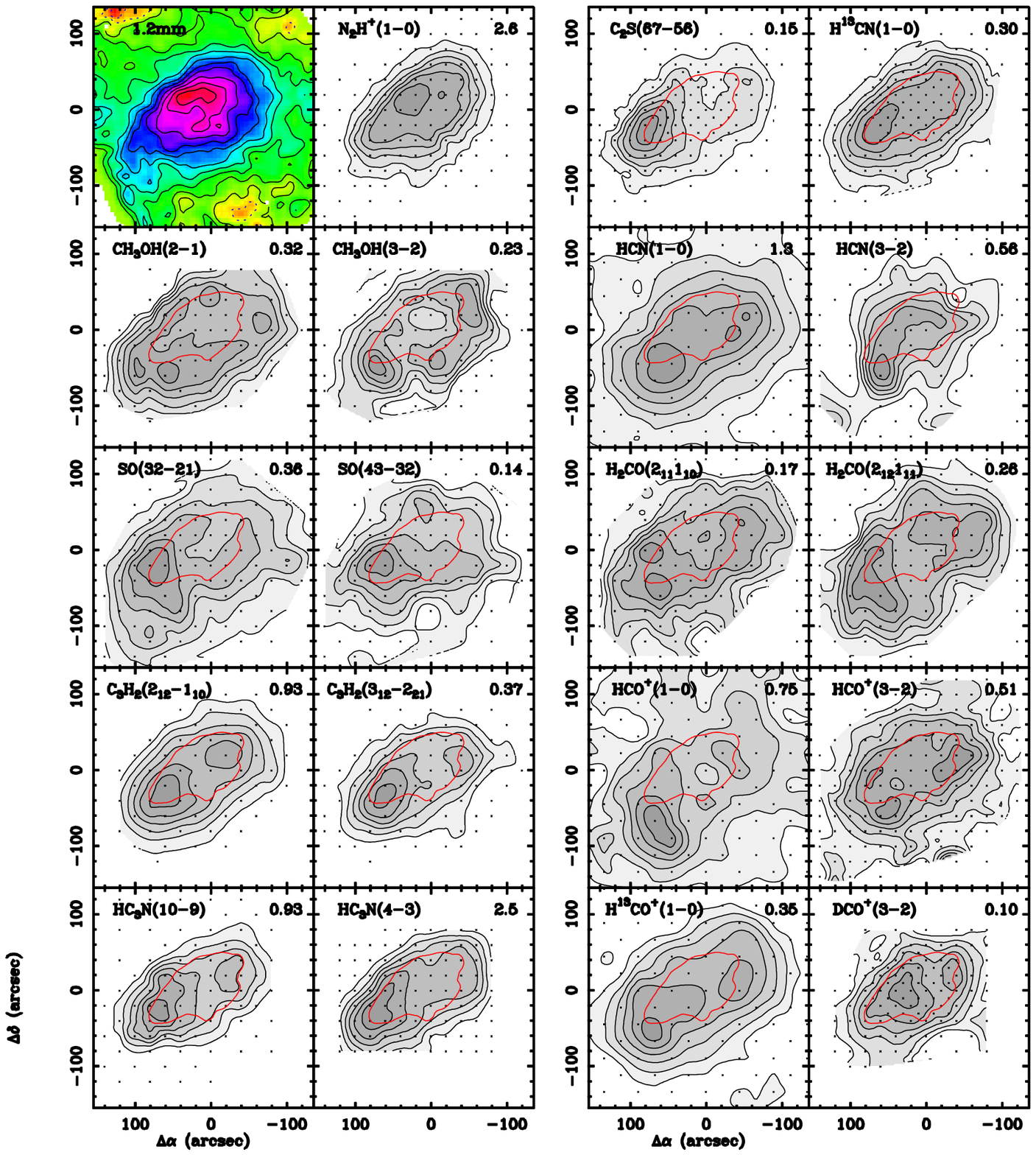}}
\caption{Integrated intensity maps from our molecular survey 
of L1498. The two top left panels show the 1.2 millimeter
continuum and N$_2$H$^+$(1--0) data presented in paper 1.
In contrast with these centrally concentrated distributions, 
the other molecular-line
maps are ring-like or centrally-depressed, and have relative
minima at the 1.2mm/N$_2$H$^+$ peak. This is an indication
of a central drop in the abundance of most species (see
text). Both IRAM 30m and FCRAO maps
have been convolved with a gaussian beam of 20-30 arcsec
to filter out high-frequency noise, and for all lines, the 
contours are at 15, 30, ... percent of the peak intensity
(peak intensity in K km s$^{-1}$ is indicated on the top right corner of
each panel). To better compare the survey line maps with the 
N$_2$H$^+$(1--0) emission, the contour at 75\% of the peak
is superimposed in red.
The dots indicate the original map sampling.
Central coordinates are $\alpha_{1950}=4^{\rm h}7^{\rm m}50\fs0$, 
$\delta_{1950}=25\degr02'13\farcs0$.
\label{fig1}}
\end{figure*}

\begin{figure*}[t]
\centering
\resizebox{15cm}{!}{\includegraphics{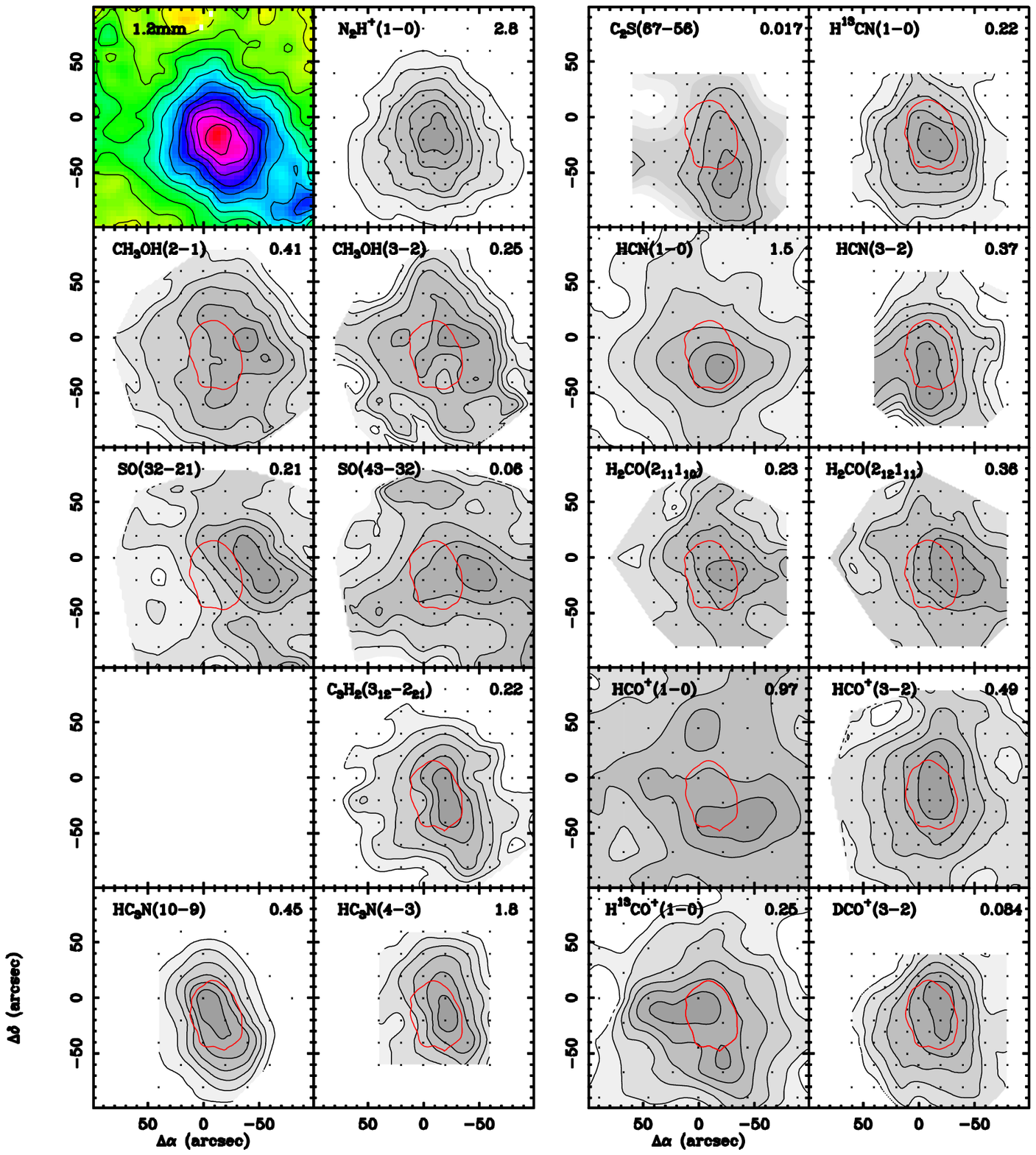}}
\caption{
Same as Fig 1 but for the L1517B core.
Central coordinates are $\alpha_{1950}=4^{\rm h}52^{\rm m}7\fs2$,
$\delta_{1950}=30\degr33'18\farcs0$.
\label{fig2}}
\end{figure*}

A simple inspection of the maps in Figs. 1 and 2 shows that 
different molecular species have depletion holes
of different sizes. Using again the larger L1498 core as a reference,
we note that CH$_3$OH presents a rather prominent 
hole, while DCO$^+$ has a central depression 
barely resolved by our observations. This variety of
hole sizes most likely arises from the different sensitivity
to depletion of the different species and from the
complex chemical changes resulting from 
the freeze out of CO, the main source of carbon in the gas phase.
The freeze out of CO decreases the abundance of the different C-bearing 
species by different degree, while it enhances the abundance of
deuterium-bearing species \citep{dal84}.  In Section 5.2 we 
will present a quantitative comparison of hole sizes 
using the results of a radiative transfer analysis.

Given the large effect of depletion in the
maps of Figs. 1 and 2, it seems clear that
in most cases the shape of the emission
reflects more the chemical composition of a
core than its true physical structure. As chemical inhomogeneities
seem the rule in starless cores, this high sensitivity to chemistry
of the maps should be carefully considered
when deriving physical properties from observations of
line emission. The risk of overlooking it  
is graphically illustrated by the maps of L1498 in species like
CH$_3$OH, SO, or C$_3$H$_2$. These maps show highly 
fragmented distributions with multiple peaks along an approximately
elliptical shell, and a naive interpretation of the emission peaks
as distinct physical structures would lead to a picture of 
a highly clumped core. This is of course in contradiction with the 
the distribution of the most reliable
tracers (1.2mm dust continuum, N$_2$H$^+$, and NH$_3$), which 
shows that the core is smooth and centrally concentrated. Any
correspondence between map peaks and core substructure,
therefore, requires careful consideration core chemistry
and self consistency check using multiple species.
The traditional warning against the use of optically thick
tracers to infer physical properties of cores should now 
be expanded to avoid using depletion-sensitive species 
for the same purpose.
Comparing in Figs. 1 and 2 the similar maps of thick and thin 
tracers like HCO$^+$ and H$^{13}$CO$^+$ (also HCN and H$^{13}$CN)
with the distinct distributions of N$_2$H$^+$ and the dust continuum,
we see that the danger of using depletion-sensitive molecules
can sometimes exceed that of using thick or even self-absorbed tracers.

\section{Abundance analysis}

To improve the qualitative abundance analysis of the previous
section we need to model the emission of the observed
species. Modeling this emission requires first determining
the physical structure of each core and then following 
the transfer of radiation. For the first step, we make
a physical model of the core that describes
its density, temperature, and gas motions, and we do so
assuming spherical symmetry because of 
the close-to-round shape of the continuum maps.
Once the core is modeled, it becomes like 
a laboratory of known physical conditions
where the line intensities can be inverted into
abundance estimates. For this second step, we use 
a non-LTE Monte Carlo code that solves numerically 
the transfer of the line emission. The details
of these two steps are described in the next section.

\subsection{Core physical model and Monte Carlo radiation transfer}

The physical models of L1498 and L1517B were derived
in paper 1 from the simultaneous fit of the dust continuum,
C$^{18}$O/C$^{17}$O, CS/C$^{34}$S, N$_2$H$^+$, and NH$_3$ emissions.
These data constrain the core density, temperature, and 
kinematics, and here we use the same parameterization for
consistency. As mentioned in paper 1, we search for
the simplest expressions consistent with the data. 
We select the core center from the continuum emission,
and fit the radial profile of this emission assuming 
a density profile of the form
$$n(r) = {n_0 \over 1+(r/r_0)^\alpha},$$
where $n_0$, $r_0$, and $\alpha$ are the central density, half-maximum
radius, and asymptotic power law. 
In this way, we estimate central densities
of $0.94 \; 10^5$ and $2.2 \; 10^5$ cm$^{-3}$ for L1498
and L1517B, respectively.
(Recent density determinations for these
two cores by \citealt{kir05} using SCUBA data seem to confirm our
estimates, although \citealt{shi05} have derived also from SCUBA data
a central density for L1498 that is more than 3 times lower than ours.)
As the cores
are embedded in clouds, we truncate the analytic density profile at
$r=4 \; 10^{17}$ cm ($190''$ at 140 pc) and continue it with a constant
density envelope of $10^3$ cm$^{-3}$ for another
$r=4 \; 10^{17}$ cm. This extension only affects
the modelling of self absorbed lines and therefore has no
influence on our abundance profiles, which we model with
thin species. The temperature profile of the cores
is described with a constant
value of 10 K for L1498 and 9.5 K for L1517B 
(from NH$_3$ emission data), and the non-thermal linewidth is
also taken as constant in the inner core, with a FWHM of 0.125
km s$^{-1}$ for both cores (a larger value is assumed in the extended
cloud, see paper 1).
Finally, the two cores are assumed to be static close to the
center ($r<1.75 \; 10^{17}$ cm for L1498 and $r<1.5 \; 10^{17}$ cm
for L1517B) and have a slow gradient outside (inward for L1498
and outward for L1517B). These outer motions are subsonic and
most likely represent the
kinematics of the extended cloud (e.g., Appendix B).

To solve the radiative transfer inside each core we also
assume spherical symmetry. 
In paper 1 we saw that the maps of C$^{18}$O and CS are not
circularly symmetric, in contrast with the maps of more reliable
tracers like the mm continuum, N$_2$H$^+$, and NH$_3$. Figs. 1
and 2 show now a similar situation for the species of our survey:
in L1498 the emission is systematically brighter towards
the SE (some species present a secondary maximum toward the
NW), and in L1517B the western half of the core is brighter than
the eastern half. These distributions suggest that the
abundance of most species is not spherically symmetric, 
despite the symmetric gas distribution inferred
for the two cores.  As studied in paper 1, the deviations
seem correlated with the velocity of the gas, and this can be understood
as the result of differential depletion caused by a
non spherical contraction of the cores.\footnote{Note that 
the caption of Fig. 13 in paper 1 erroneously labels the N$_2$H$^+$ 
emission in L1498 as red when is blue.} Modeling these
asymmetric distributions requires a 2D or even 3D radiative transfer
code, which exceeds the scope of this paper. In the following 
discussion we assume spherical symmetry and fit for each species 
a circular average of the emission. In this way, our abundance 
estimates represent azimuthal averages over the core and therefore 
correspond to the result of a mean contraction.

As in paper 1, we solve the radiative transfer inside each core 
with the non-LTE Monte Carlo code of \citet{ber79}, which we
have modified to include additional molecules (see
Appendix A for a summary of the molecular parameters
used in this work). For each molecular species, we run the code
together with the core physical model to produce
a set of expected intensity distributions for different abundance
profiles. These intensity predictions are convolved with the appropriate
gaussian beam and compared with the observed
radial profile of integrated intensity
and the central emerging spectrum for as many transitions
of the species as we have observed (usually two); the best fit model
is the one that fits all these constraints simultaneously. As
the only free parameter in this process is the
abundance profile, observations of one transition are in 
principle enough to constrain the solution. Fitting 
at least two transitions simultaneously, as we do, checks the
self-consistency of the process.

\begin{figure*}[th!]
\centering
\resizebox{14cm}{!}{\includegraphics{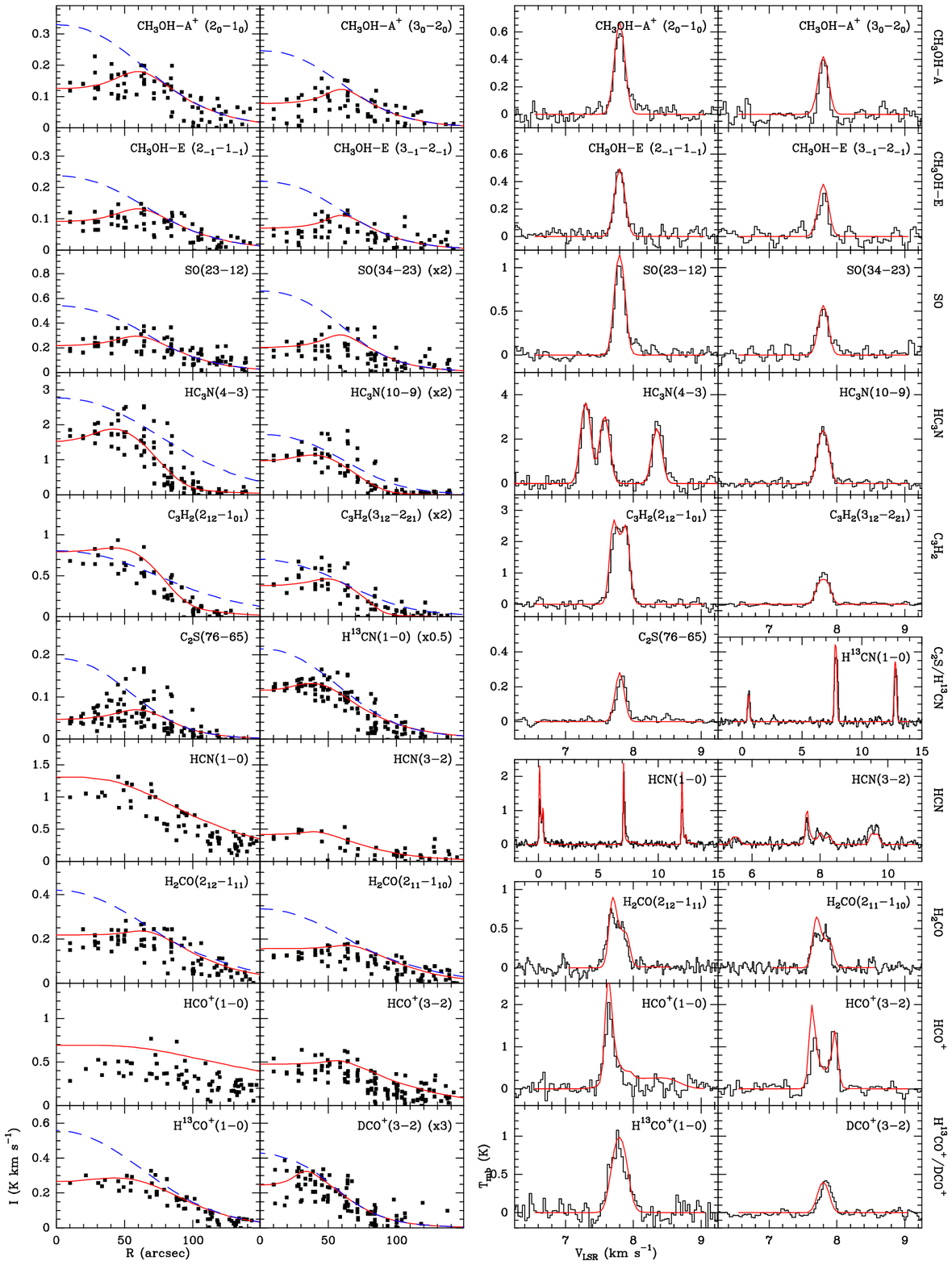}}
\caption{Radial profiles of integrated intensity and central
spectra for all line data observed toward L1498. The squares
in the radial profiles and the histograms in the spectra 
are observed data while the lines represent Monte Carlo model
results. The blue dashed lines are models that assume a
constant abundance chosen to fit the data at a relatively
large fiducial radius (see text), and the red solid lines are
the prediction from the best fit model. Note how the constant 
abundance models over predict the central intensity typically
by a factor of 2 or more.
The integrated intensities in the radial profiles (left panels) have
a typical 1-$\sigma$ uncertainty of 0.01-0.02 K km s$^{-1}$ for the 
IRAM data (with the exception of the high frequency lines
HCN(3--2) and HCO$^+$(3--2), which have 1-$\sigma$ uncertainties
of 0.06 and 0.09 K km s$^{-1}$), 0.02-0.065 K km s$^{-1}$ for the
FCRAO data, and 0.09 K km s$^{-1}$ for the 100m data. These values
are smaller than the internal point-to-point scatter.
The spectra in the right panels have been generated from the average 
of typically 5 spectra within a $20''$ radius from the core
center ($50''$ for FCRAO data), in order to produce a relatively 
high S/N observation. To model this observation, we have generated
synthetic spectra for each of the (typically 5) observed radial
locations, and we have averaged them as we have done with the data.
\label{fig3}}
\end{figure*}

\begin{figure*}[t]
\centering
\resizebox{15cm}{!}{\includegraphics{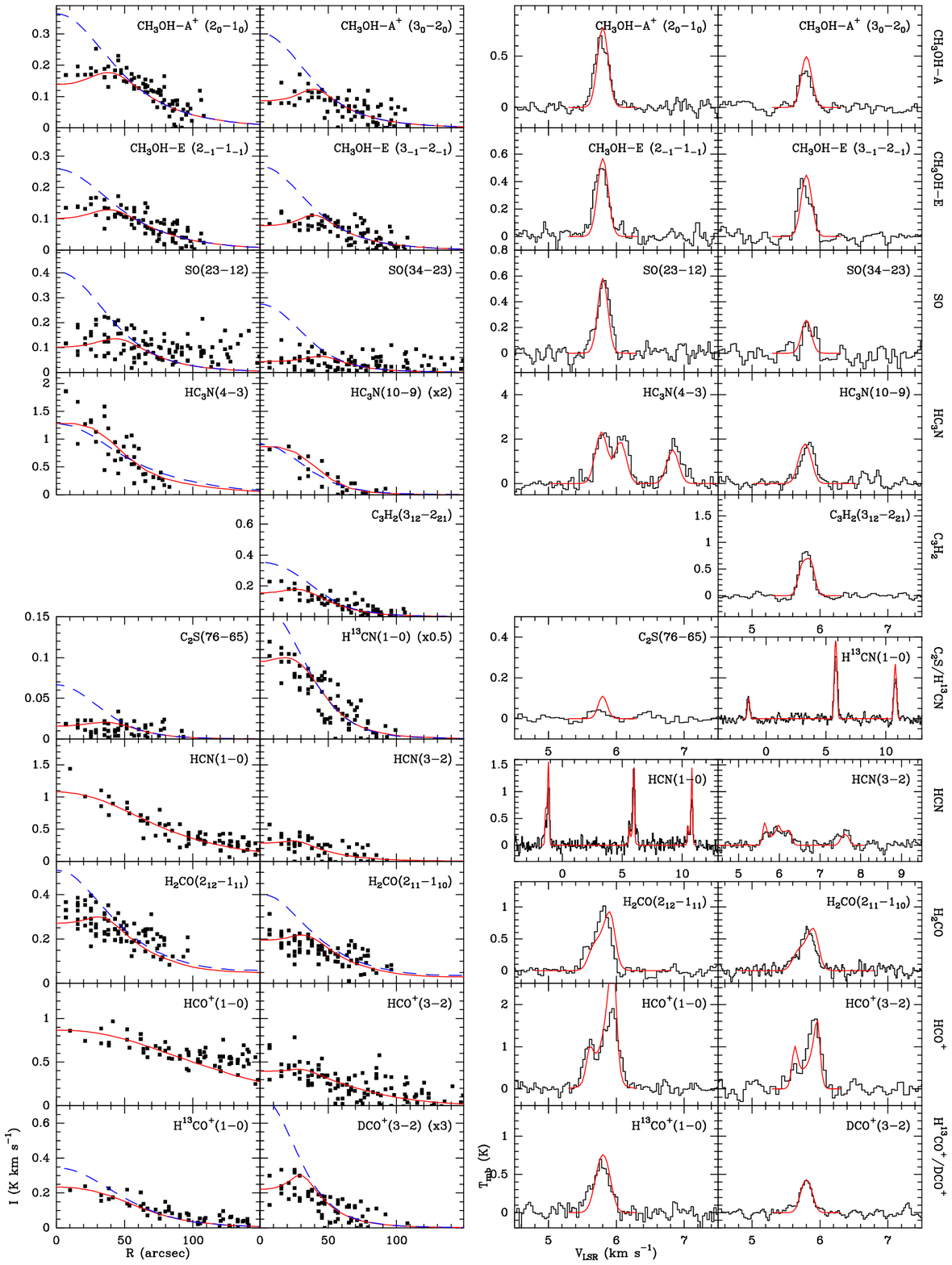}}
\caption{Same as Fig. 3 but for the L1517B core.
\label{fig4}}
\end{figure*}

\subsection{Fitting procedure}

As when fitting the core physical parameters, we aim for
the simplest abundance profiles consistent with the data. For  
each species, we start with a constant abundance model that
fits the integrated intensity at a fiducial outer radius.
This radius is chosen as $75''$ for L1498 and $55''$ for L1517B
and represents a compromise between the need of a large enough radius 
(to detect central molecular depletion) and the need of 
a bright enough signal
(to make a reliable fit). As shown by the dashed lines in 
Figures 3 and 4, the constant abundance models systematically
overestimate the intensity toward the core center.

To improve the fit, we decrease the abundance toward the core center
by introducing a step-function at $r_h$. 
The abundance inside $r_h$ is taken to be
negligible ($10^{-4}$ of the outer abundance), and the value of $r_h$
is used as a free parameter to improve the quality of the fit. 
This is the same approach used in paper 1 to derive abundance 
profiles for C$^{18}$O and CS, and for most species, 
it provides a reasonable fit.
In a few cases, a slightly different parameterization
is needed, and these exceptions are detailed below.

Parameterizing the abundance profiles with a step function
is clearly a simplification of the expected
depletion behavior of molecules under realistic core
conditions (e.g., \citealt{ber97,aik05}).
However, given the fast abundance drops predicted by these models
and the finite spatial resolution of our observations, 
this parameterization is almost as good a description, and
it has the advantage of allowing an easy comparison between 
molecules by comparing the outer abundance and $r_h$ values of 
their fits. In
section 5  we will use this approach to study different aspects
of the behavior of molecules under high density conditions.
In the following paragraphs, we present further details on
the fitting procedure of each individual species. Readers not interested
in such a detailed view can simply inspect Figures 3 and 4 and 
move to section 5.

{\em CH$_{\mathit3}\!$OH.} Two transitions were observed 
for each of the A and E forms 
of this molecule, so we have fitted the two forms independently and 
determined 
E/A ratios of 1 for both L1498 and L1517B. As Figs. 3 and 4 show,
the constant CH$_3$OH abundance models fail to fit the central intensity
by about a factor of 2, while the models with
a central hole provide reasonable fits to all data simultaneously.

{\em SO.} SO is the only non C-bearing molecule observed in 
this survey (N$_2$H$^+$ and NH$_3$ were studied in paper 1), 
so its behavior provides information complementary to that of
all other species.
Three SO transitions were observed toward L1498 and L1517B, 
but the SO(JN=12--21) line 
was not detected in either core. Our fits therefore only take into account 
the emission from SO(32--21) and SO(43--32),
as the 12--21 non detection is trivially fitted by all models.
Again, the constant abundance models fail to
fit the observed SO intensity toward the core centers by
a wide margin. Models with a central
hole, on the other hand, reproduce the intensities both at large 
and small radius, and show that this species is one of the most sensitive
indicators of molecular depletion (section 5.5). 

{\em HC$_{\mathit3}\!$N.} 
In L1498, the model with constant HC$_3$N abundance
does not reproduce the observed combination of compact emission
and a central hole. Such a pattern requires that the HC$_3$N abundance 
increases inwards at intermediate radii, and that it has a sharp
depletion hole near the core center. We fit this 
behavior by introducing an inward jump in the abundance by a
factor of 10 at $r = 1.8 \; 10^{17}$ cm ($85''$ at 140 pc) 
together with the step central hole
used for the other species. This type of profile
fits well both the J=4--3 and 10--9 data (Fig. 3), together with 
the central 12-11 spectrum (not shown). It also agrees with
the prediction of the chemical model by \citet{ruf97}, who
find a late-time HC$_3$N peak caused by CO depletion.
In L1517B the situation is
less clear due to the smaller size of this core. A constant
abundance model barely fits the data and is inconsistent
with the lopsided emission seen in the 4--3 
map. To improve the fit, we have introduced a
central hole with a small radius ($r = 4 \; 10^{16}$ cm =  $20''$)
and a factor of 2 outer abundance drop at $r = 1.25 \; 10^{17}$ cm
($60''$). This small inner hole, at the limit of our resolution, is
consistent with the smaller-than-average hole found in L1498. As
will be further discussed in section 5.2, HC$_3$N seems to survive
in the gas phase to higher densities than other species.

{\em C$_{\mathit3}$H$_{\mathit2}$.}
In L1498, C$_3$H$_2$ presents the same combination of
compact emission and central hole seen in HC$_3$N. 
A constant abundance model that reproduces the emission 
at intermediate
radii fails to fit the observations both at large and small
radii (Fig. 3). As with HC$_3$N, we introduce a factor of 
10 inward jump in the abundance at $r = 1.8 \; 10^{17}$ cm 
together with a central hole. This model 
fits well the emission, including the
slightly self-absorbed 2$_{12}$--1$_{01}$
spectrum towards the core center (Fig. 3).
We note that although there are no theoretical 
predictions for a
late-time enhancement of C$_3$H$_2$ (\citealt{ruf97} do not
present results for this species), the chemistry
of C$_3$H$_2$ and HC$_3$N seem related (e.g., 
\citealt{cox89}). It is therefore possible that a single
process explains the observed behavior of the C$_3$H$_2$ and 
HC$_3$N in L1498.
Unfortunately, only one C$_3$H$_2$ transition was observed
toward L1517B. Lacking the brighter 2$_{12}$--1$_{01}$
transition, we have fitted the C$_3$H$_2$ abundance in this core
with only a central hole and no abundance jump 
at intermediate radii. Observations of additional lines
of this molecule are needed to clarify the behavior of
C$_3$H$_2$ in L1517B.

{\em C$_{\mathit2}$S.} Only the JN=67--56 transition was observed toward
both cores and, as Figures 3 and 4 show, its radial
distribution is inconsistent with a constant abundance profile.
A model with a central abundance hole fits the emission
both at large and small radii in both cores and shows that, like SO,
C$_2$S is highly sensitive to depletion. Our model for
L1498 also fits nicely the emission of the JN=21--10 (22 GHz),
43--32 (45 GHz), and 87--76 (94 GHz) lines observed
by \citet{wol97}, who with \citet{kui96}, first 
found central depletion for this molecule.
A detailed comparison with these published
data is presented in Appendix C. 

{\em HCN and H$^{\mathit{13}}\!$CN.} The HCN(1--0) spectra show evidence for
self absorption towards both L1498 and L1517B (see also \citealt{soh04}).
The HCN emission is therefore dominated by
the core outer layers and it cannot be used to derive 
an abundance profile inside the core. We thus base our abundance
determination on the thinner H$^{13}$CN(1--0) emission, which
clearly shows a need for a central abundance drop in both cores. 
As in paper 1, we assume 
a $^{12}$C/$^{13}$C ratio of 60, and we model the main isotope emission by
scaling the H$^{13}$CN abundance profile by that factor. This produces
a fit of the HCN(3--2) emission in both cores but 
underestimates 
slightly the depth of the 1--0 self absorptions. To improve
the fit, we introduce in the constant density envelope that surrounds each 
core ($r > 4 \; 10^{17}$ cm) an abundance enhancement
of a factor of 4 in L1498 and a factor of 2 in L1517B (the
enhancement has no effect on the H$^{13}$CN emission or even on the 3--2 line).
It should be noted that $r > 4 \; 10^{17}$ cm
corresponds to radii larger than $190''$, which are outside
the region where our dust-based determination of the density is
reliable. For that reason, the outer abundance enhancements
should be considered a fitting convenience that corrects the 
simple core+envelope model that we have used in the Monte
Carlo radiative transfer, and not necessarily an indication of an abundance
change in the outer core. It is in fact remarkable that such a simple
parameterization can simultaneously fit both the radial profiles
and the central spectra of lines with such different optical 
depths as H$^{13}$CN(1--0), HCN(1--0), and HCN(3--2) (Figs. 3 and 4).

{\em H$_{\mathit2}\!$CO.} Our two H$_2$CO lines, 
2$_{12}$--1$_{11}$ and 2$_{11}$--1$_{10}$, show evidence for self-absorption 
and, unfortunately, no thin isotopologue of this species was observed in
the survey. Despite this, constant abundance models can be easily ruled 
out from the shape of the radial profiles,
and a central abundance drop is needed to fit simultaneously
the inner and outer emission. To properly
model the self absorption, we need again an abundance
increase in the envelope ($r > 190''$, and factors 15 and 5 for L1498 and 
L1517B, respectively), to which the same caveats mentioned for HCN apply. 
\citet{you04} have observed two additional
lines towards L1498 (1$_{11}$--1$_{10}$ and 3$_{12}$--2$_{11}$) 
and have also concluded
that a central abundance drop is needed to fit the data.
As shown in Appendix C, our best fit model also fits the
H$_2$CO(3$_{12}$--2$_{11}$) data from \citet{you04}, 
while it predicts a 1$_{11}$--1$_{10}$
self-absorption that is weaker than observed.
This 1$_{11}$--1$_{10}$ absorption originates at the core outer
envelope (Appendix C), so the failure in the fit
results again from our simplified parameterization of the
core outer layers. 

{\em HCO$^+$, H$^{\mathit{13}}\!$CO$^+$.} Both the J=1--0 and 3--2
transitions of HCO$^+$ are deeply self absorbed, so our abundance
determination relies on the thinner H$^{13}$CO$^+$(1--0) line 
and assumes again a $^{12}$C/$^{13}$C ratio of 60. 
As in all other species, a central drop in the HCO$^+$
abundance is needed to fit the shape of the radial profiles.
To fit the HCO$^+$ self absorption, we again 
need to include an outer envelope ($r> 190''$)
abundance enhancement of a factor of 4 in the L1498 model
(still the predicted 1--0 line is
brighter than observed mostly due to poor modelling 
of the red component) while  
no abundance enhancement is needed for L1517B.

{\em DCO$^+$.}
A central abundance hole is also required to fit the DCO$^+$ data, 
but its radius is smaller than the HCO$^+$ hole radius. This difference
most likely results from a central increase in the deuterium fractionation 
caused by the CO depletion (e.g., \citealt{cas99}), which partly compensates 
the DCO$^+$ freeze out at the inner core. The effect is of course 
incomplete because the DCO$^+$ abundance end ups falling at the 
highest densities. Still, the presence of DCO$^+$ inside the
CO and HCO$^+$ depletion holes suggests that a
small amount of CO and HCO$^+$ survives at high densities.
Comparing the DCO$^+$ and HCO$^+$ abundances in 
the outer core, we derive deuteration ratios
of 0.02 and 0.03 in the outer layers of L1498 and L1517B,
respectively. Each of these values is a factor of 2 lower
than the deuteration ratios measured toward the inner core
by \citet{cra05} using N$_2$H$^+$ and N$_2$D$^+$, in good agreement
with the expectation of a degree of deuteration that increases 
toward the core center.

Although the abundance enhancements at large radii required
to fit the self absorptions in HCN, H$_2$CO, and HCO$^+$
need confirmation using a model of the large-scale structure
of the cores, we note that such enhancements are not totally
unexpected. Absorption line studies by \citet{luc96} and
\citet{lis01}
reveal that species like H$_2$CO, HCO$^+$, and HCN present 
large column densities in diffuse clouds (but not N$_2$H$^+$, 
for example). It is therefore
possible that the lower density gas surrounding
L1498 and L1517B has a chemical composition
similar to the diffuse clouds studied by Liszt \& Lucas.
This would naturally explain the large amounts of 
low-excitation gas needed to produce the observed self absorptions.

\subsection{Ortho-to-para and isotopic ratios: total
abundances}

For species like NH$_3$, H$_2$CO, and C$_3$H$_2$, our observations 
and modeling only determine the abundance of the ortho or para 
form of the molecule, and we need to assume an ortho-to-para ratio (OPR)
to convert our estimate into a total molecular abundance.
This OPR depends on the formation history of 
the molecule, and is therefore somewhat uncertain. Here we assume that the
three species are formed by gas-phase reactions, because grain production
will require a mechanism to release the products to the gas
phase, a difficult task given the strong freeze out observed.
(This of course does not imply that NH$_3$, H$_2$CO, and C$_3$H$_2$ do not form 
on dust grains, but that the 
gas-phase chemistry is separate from the dust-grain production,
as seen in the models of, e.g., \citealt{aik05}.) If this is the case, 
the OPR should be close to 
the high-temperature limit, because the energy difference
between ortho and para forms (2.4 K for C$_3$H$_2$, 15.2 K for H$_2$CO, and 
23.4 K for
NH$_3$) is significantly lower than the energy released during molecule
formation (e.g., \citealt{dic99}). Indeed, OPRs close to 3 (high
temperature limit) have been found for H$_2$CO in starless cores, including
L1498, by \citet{dic99}. \citet{tak01} measure
a C$_3$H$_2$ OPR of 2.4 towards TMC1, also close to the high temperature 
limit. We thus assume OPRs of 3 for H$_2$CO and C$_3$H$_2$ and 1 for NH$_3$.

In addition to OPRs, we
need to assume isotopic ratios to convert rare isotopologue abundances into
main species values. Along this paper and in paper 1 we have used the
standard isotopic ratios
of $^{12}$C/$^{13}$C = 60, $^{32}$S/$^{34}$S = 22, and 
$^{18}$O/$^{17}$O = 3.65, to which we now add the terrestrial 
$^{16}$O/$^{18}$O ratio of 500. Using these values, we estimate the
final main isotopologue abundances
presented in Table 2. In the following
discussion we will refer only to the main molecular 
forms, although it should be remembered that most abundances are
based on the rare (and optically thin) isotopologues.

\begin{table}
\caption[]{Molecular abundances with respect to H$_2$
\label{tbl-2}}
\[
\begin{array}{lrcrcc}
\hline
\noalign{\smallskip}
\mbox{} & \multicolumn{2}{c}{\mbox{L1498}} 
& \multicolumn{2}{c}{\mbox{L1517B}} & \\
\mbox{Molec.} & X_0{\mbox{~~~~~}} & r_h & X_0{\mbox{~~~~~}} & r_h & \mbox{Notes} \\
\mbox{} & & \mbox{(10$^{17}$ cm)} &  & \mbox{(10$^{17}$ cm)} & \\
\noalign{\smallskip}
\hline
\noalign{\smallskip}
\mbox{CO} & 2.5 \; 10^{-5} & 1.5 & 7.5 \; 10^{-5}  & 1.75 & \\
\mbox{CS} & 3.0 \; 10^{-9} & 1.0 & 3.0 \; 10^{-9}  & 1.15 & \\
\mbox{N$_2$H$^+$} & 1.7 \; 10^{-10} & 0 & 1.5 \; 10^{-10}  & 0 & (1)\\
\mbox{NH$_3$} & 2.8 \; 10^{-8} & \beta=3 & 3.4 \; 10^{-8}  & \beta=1 & (2) \\
\mbox{CH$_3$OH~~~} & 6.0 \; 10^{-10} & 1.2 & 6.0 \; 10^{-10}  & 0.8 & (3) \\
\mbox{SO} & 4.0 \; 10^{-10} & 1.2 & 2.0 \; 10^{-10}  & 0.9 \\
\mbox{C$_3$H$_2$} & 1.6 \; 10^{-9} & 1.0 & 9.3 \; 10^{-10}  & 0.6 & (4) \\
\mbox{HC$_3$N} & 1.0 \; 10^{-9} & 0.85 & 4.5 \; 10^{-10}  & 0.4 & (5) \\
\mbox{H$_2$CO} & 1.3 \; 10^{-9} & 1.3 & 6.7 \; 10^{-10}  & 0.7 & (6) \\
\mbox{HCO$^+$} & 3.0 \; 10^{-9} & 1.15 & 1.5 \; 10^{-9}  & 0.6 & (7) \\
\mbox{DCO$^+$} & 5.0 \; 10^{-11} & 0.65 & 5.0 \; 10^{-11}  & 0.6 \\
\mbox{C$_2$S} & 4.0 \; 10^{-10} & 1.25 & 1.0 \; 10^{-10}  & 0.8 \\
\mbox{HCN} & 7.0 \; 10^{-9} & 0.8 & 3.0 \; 10^{-9}  & 0.53 & (8) \\
\noalign{\smallskip}
\hline
\end{array}
\]
\begin{list}{}{}
 \item[$(1)$] Factor of 3 drop for $r > 1.8 \times 10^{17}$ cm in L1498
 \item[$(2)$] OPR = 1 assumed, $X$(NH$_3$) = $X_0 \; (n(r)/n_0)^\beta$
 \item[$(3)$] E/A = 1
 \item[$(4)$] OPR = 3 assumed and factor of 10 drop
 for $r > 1.8 \times 10^{17}$ cm in L1498
 \item[$(5)$] Factors of 10 (L1498) and 2 (L1517B) drops for 
 $r > 1.8 \times 10^{17}$ cm
 \item[$(6)$] OPR = 3 assumed. Factors of 15 (L1498) and 5 (L1517B) enhancement 
 for $r > 4 \times 10^{17}$ cm
 \item[$(7)$] Factor of 4 enhancement for $r > 4 \times 10^{17}$ cm in L1498
 \item[$(8)$] Factors of 4 (L1498) and 2 (L1517B) enhancement 
 for $r > 4 \times 10^{17}$ cm
 \end{list}
 \end{table}

Finally, we stress the dependence of our abundance
estimates on the assumed dust parameters. As mentioned in 
paper 1, the values of the dust temperature and emissivity
are still poorly known (e.g., \citealt{bia03,kra03}), and 
their uncertainty propagates
into the determination of the gas density profile and 
total core column density. These determinations, in turn, affect
the abundance estimates through their combined effect on the 
level excitation and the H$_2$ column density. For most species,
the abundance estimate uses
optically thin lines from low-energy levels, so the effect of
the density on the excitation is smaller than its effect
on the column density. Thus, to first order, we can
correct for dust parameter changes with the following simple equation:
$$X(\kappa, T_d) = X(0.005,10)\; \left({\kappa_{1.2mm} \over 0.005 
\mathrm{~g~cm}^{-2}}\right) \; 
\left({B_\nu(T_d) \over B_\nu(10 \mathrm{K})}\right),$$
where $B_\nu$ is the Planck function,  
$X(0.005,10)$ is our abundance determination using a 1.2mm dust
emissivity $\kappa_{1.2mm} = 0.005$ cm$^2$ g$^{-1}$ and a dust
temperature of $T_d = 10$~K, and $X(\kappa, T_d)$ is the
corresponding abundance for arbitrary values of $\kappa_{1.2mm}$
and $T_d$.  If, for example, $\kappa_{1.2mm} = 0.01$ cm$^2$ g$^{-1}$
\citep{oss94}
and $T_d$ = 8 K \citep{eva01, gal02}, our abundance values will
need to be multiplied by 1.4.

\section{Discussion}

\subsection{Comparison between L1498 and L1517B}

\begin{figure}[t]
\centering
\resizebox{\hsize}{!}{\includegraphics{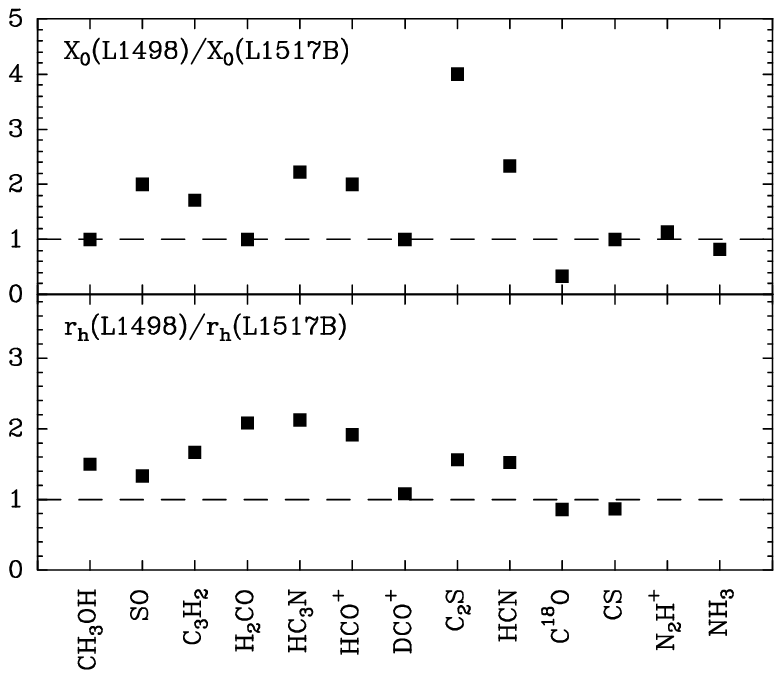}}
\caption{ Abundance comparison between L1498 and L1517B. {\em Top:}
ratio between the abundances outside the central hole
($X_0$) in the two cores
(for HC$_3$N and C$_3$H$_2$, X$_0$ is the abundance between the
hole and the outer cutoff). For most species, this ratio
is close to 1 (within a factor of 2), indicating a similar
pre-freeze out composition. {\em Bottom:} ratio between
central depletion holes ($r_h$). L1498 presents systematically larger
hole ratios by a factor of about 1.5, which is consistent  with the
the visual impression from the integrated maps.
\label{fig5}}
\end{figure}

L1498 and L1517B present similarities and differences in their
physical properties. They have almost the same kinetic temperature,
level of turbulence, and total gas column density, while they
differ by a factor of 2 in central density and half
maximum density radius (paper 1). Analogously,
the two cores present similarities and differences in their chemical
composition. To examine them here, we compare the $X_0$ and $r_h$
values derived for the different species.

The top panel of Figure 5 shows the ratio between the outer molecular
abundances derived in L1498 and L1517B
for each molecule in our survey (values for C$^{18}$O, CS, N$_2$H$^+$, and
NH$_3$ are from paper 1).
As can be seen, there is good agreement between the $X_0$ parameters 
estimated for the two cores, with most differences being smaller than a 
factor of 2. The main outliers in the plot 
are C$_2$S (4 times more abundant in
L1498 than in L1517B) and C$^{18}$O (3 times more abundant in L1517B),
and although
their abundance differences could be real, it should be noted that 
their $X_0$ determination is specially prone to error: C$_2$S is 
depleted up to such a large radius in L1498 that $X_0$ is determined using
very limited information of the outer core, and the $X_0$ 
value for C$^{18}$O can be contaminated by the extended emission
from the ambient cloud. Given these and other uncertainties, it is rather
remarkable that most $X_0$ values agree within less than a factor
of two, and we take this fact as an indication that the two cores
share very similar molecular compositions in their outer (undepleted)
layers.

The bottom panel of Fig. 5 presents a
comparison between the radius of the central abundance holes 
in L1498 and L1517B. In addition to some scatter, the figure reveals
a trend for $r_h$ to be larger in L1498 by a factor of about 1.5.
This trend agrees with the impression from the maps of Figs. 1 and 2
that L1498 has a larger depletion hole than L1517B for most
or all molecular species. Unfortunately, 
the scatter and the uncertainties in the analysis
make it impossible to determine
whether the $r_h$ ratio varies with molecule. 

The similar outer abundances of L1498 and L1517B 
suggest that both cores have contracted from ambient gas
with similar initial compositions. It is unclear however
whether the different $r_h$ values result from the 
cores being at different evolutionary stages or from
them having followed different contraction paths.
L1517B is more centrally peaked, denser at the center, and more circularly 
symmetric than L1498, so it would seem to be at a more 
advanced evolutionary stage. If this is the case,
and both cores are following the same contraction path,
the smaller $r_h$ of L1517B would imply that the depletion hole 
in a core shrinks as it contrasts; this could result from 
the contraction time scale being shorter than 
the freeze out time scale during the late stages of contraction.
Alternatively, L1517B may have
contracted faster than L1498, so its central gas has had less
time to freeze out. Further studies of core composition using a
larger sample of systems are needed to clarify this important
issue.

\subsection{Comparison between molecules}

\begin{figure}[t]
\centering
\resizebox{\hsize}{!}{\includegraphics{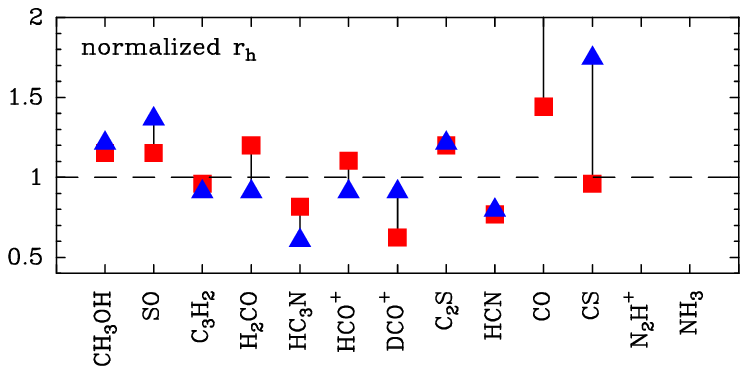}}
\caption{Normalized radius of the abundance hole $r_h$ for all molecules
in the survey. The red squares represent L1498 data and the blue
triangles are L1517B data (normalizing radius is $1.04 \times 10^{17}$ cm
for L1498 and $0.66 \times 10^{17}$ for L1517B). Species like CO, SO,
and C$_2$S
present larger $r_h$ values, while 
HC$_3$N, DCO$^+$, and HCN have smaller $r_h$ values and therefore
must remain in the gas phase to higher densities.
\label{fig6}}
\end{figure}

The radiative transfer results of
Table 2 suggest that different molecules
deplete at different radii. In addition
to the extreme behavior of N$_2$H$^+$ and NH$_3$, which do not
present abundance holes, several molecules 
have holes that are systematically smaller or 
larger than average in both L1498 and L1517B.
Identifying these molecules can help better understand
the details of depletion. It can also provide
a set of tracers to selectively sample cores at 
different depths. 

To compensate for the systematically larger
holes in L1498, we normalize the hole radii. We do this
by dividing the radius of each species by the average hole radius 
in the core, which is calculated from the mean radius of all
species with small dispersion (i.e.,
we exclude CO and CS, see below, in addition to 
N$_2$H$^+$ and NH$_3$). This normalized hole radius is 
presented in Fig. 6 for all species in our survey.
As the figure illustrates, the hole radii span a range of 
about a factor of 2. The CO and CS estimates 
present large scatter, which could result from 
a real difference between the cores or from artifacts like
contamination by the extended cloud in the case of CO and
the fact that L1498 the CS analysis is based on the optically 
thin C$^{34}$S emission while the analysis in L1517B used the thicker CS line
(as C$^{34}$S was too weak). All other species, on the other hand, 
present a reasonably low scatter between their L1498 and
L1517B results. This low scatter allows distinguishing between
species with relatively large and small depletion holes. In the first
group we find SO, C$_2$S, and CH$_3$OH, which in both cores have
larger-than-average abundance holes. In the second group we find
HC$_3$N, DCO$^+$, and HCN, whose central holes are more than
50\% smaller than the holes in the first group. Such a relatively
large difference in $r_h$ suggests a corresponding difference in the
depletion behavior of these species. It also shows that
HC$_3$N, DCO$^+$, and HCN survive in 
the gas phase up to higher densities than most other species,
and this makes them interesting
line tracers of the middle layers of dense cores.
Recent infall searches in starless cores \citep{lee04,soh04},
do use this property of DCO$^+$ and
HCN to penetrate deeper than using traditional tracers
like CS or HCO$^+$ (e.g., \citealt{lee01}). Our L1498 and L1517B
analysis, however, shows that HCN and DCO$^+$ 
are no substitute for N$_2$H$^+$ or NH$_3$, as they
end up depleting at densities of about $10^5$ cm$^{-3}$.

\subsection{Comparison with other starless cores}

TMC1 is the core whose chemical composition has been studied with
most detail,
and its abundances are commonly used as standards for comparison 
with other low-mass star-forming regions. Current estimates of the TMC1 
abundances, however, do not consider the effect of
molecular depletion, so a comparison with our L1498 and L1517B 
estimates can only be approximate. Until a more
realistic analysis of TMC1 is available, the most complete study 
is that of \citet{pra97}, who have presented FCRAO observations 
and analysis of more than a dozen molecular species,
8 of them common to our L1498/L1517B survey. \citet{pra97} normalize
their abundances to HCO$^+$, and provide values for three 
positions along the TMC1 filament: the cyanopolyyne peak, the ammonia peak, and 
the SO peak. To compare with these values, we select the L1498 and L1517B 
abundances outside the depletion hole and  
normalize them to HCO$^+$. The resulting values agree
well with those of TMC1, and in no case the difference exceeds
one order of magnitude. The best agreement occurs
for the cyanopolyyne peak, where the average ratio between the L1498
and TMC1 abundances for the 8 species is $1.3 \pm 1$. This position also
provides the best match to the L1517B abundances, although the ratio
is significantly larger ($2.3 \pm 2$) because 
our estimate of the HCO$^+$ in L1517B is half of that in L1498. 
Given the very different methods 
and the uncertainties in the two analysis, the agreement with
the TMC1 estimates seems rather good. The
fact that the best match to the abundance in the 
undepleted outer parts of L1498 and L1517B occurs for
the cyanopolyyne peak is also in good
agreement with the interpretation that this TMC1 position is the most 
chemically young of the filament (\citealt{suz92, hir92}).
It also suggests that the abundances of TMC1 are not anomalously high
\citep{how96},
but representative of the population of dense cores in the Taurus cloud.

A different set of abundance determinations in starless cores
has been recently provided by the Texas group using 
a technique similar to ours: 
density estimates using continuum (SCUBA) data 
followed by Monte Carlo modeling of the line emission \citep{eva01,lee03}.
As a result of this work, \citet{lee03} have estimated
the abundance of HCO$^+$, H$^{13}$CO$^+$, and DCO$^+$ in L1512, L1544,
and L1689 by fitting step functions with non zero value at the
core center. Comparing the outer abundances estimated by these authors
with those in L1498 and L1517B, we find reasonable (factor of 2) agreement
for HCO$^+$ in L1512 and L1544, the two Taurus cores, but almost 
one order of magnitude difference for H$^{13}$CO$^+$ and DCO$^+$ in the
same objects. This result is somewhat surprising, especially for 
H$^{13}$CO$^+$, as our abundance values agree with a standard isotopic
$^{12}$C/$^{13}$C ratio of 60, while the \citet{lee03} values imply a ratio
one order of magnitude lower. The origin of this
discrepancy is not clear, although it most likely
results from a combination of a higher dust emissivity, lower gas
temperature, and smaller radius coverage by \citet{lee03}. 
A similar (factor of 4) discrepancy occurs between our H$_2$CO 
abundance determination in L1498 and that by \citet{you04}
in the same core, despite our model fitting the H$_2$CO(3$_{12}$--2$_{11}$)
emission from these authors (Appendix C). This discrepancy most likely arises
from a different choice of dust emissivity, which as discussed by \citet{shi05},
makes our L1498 models differ from that of the
Texas group by a factor of 3 in H$_2$ column density. Such a systematic
discrepancy between models 
highlights the urgent need for an accurate determination of the
dust emissivity at millimeter and submillimeter wavelengths.

Better agreement between our abundance determinations and those in other 
starless cores occur for N$_2$H$^+$. \citet{cas02a} have estimated
an N$_2$H$^+$ abundance of $(2 \pm1) \times 10^{-10}$ for a sample of
25 starless cores assuming virial equilibrium, and this result is in
excellent agreement with our determination. Also, \citet{ket04}
have carried out a detailed analysis of the N$_2$H$^+$ emission
in 3 starless cores, one of them being L1517B. They derive for this object 
a central density within a factor of 2 of ours, and an
N$_2$H$^+$ abundance that differs from ours by less than 10\%.

Even if the larger abundance differences
between cores are real, it seems that in most cases 
they are well within the factor of 3-4 range. This
of course does not imply that starless cores form an homogeneous family.  
The gradient across TMC1 reveals differences in abundance that
are unlikely to be explained simply as the result of differential depletion,
and a number of cores are known to have significantly lower abundance 
of certain late-time species like NH$_3$ and N$_2$H$^+$ \citep{suz92}. L1521E, 
for example, has negligible depletion of C-bearing species and 
NH$_3$ and N$_2$H$^+$ abundances one order of magnitude lower
than L1498 and L1517B \citep{suz92,hir02,taf04b}.
Our molecular survey of the L1521E core
(in preparation) seems to suggest, however, that most of 
chemical differences between cores can be explained as a result of differential
depletion plus time evolution of the N-bearing species, and that the
undepleted abundances in the outer layers do not present extreme 
(more than a factor few) core to core
variations. A systematic survey of a larger 
sample of cores is still needed to confirm this result.

\subsection{Comparison with the Aikawa et al. (2005) chemical model}

The chemical evolution of cores as they contract under
gravity has been studied by different authors. \citet{ber97}
and \citet{cha97} were first to model the differential
depletion of C and N-bearing species as a result of 
their freeze out onto the dust grains together with
a lower binding energy of N$_2$. 
More complete chemical networks coupled to
realistic contraction physics have been used to
improve on this earlier work
\citep{aik01,li02,aik03,lee03,she03}. Very recently,
\citet{aik05} (AHRC05 hereafter) 
have presented the most up-to-date chemical 
model of starless core contraction. These authors 
have followed the evolution of 
two dense cores initially having Bonnor-Ebert density profiles, 
one subject to a small over-density factor (their $\alpha=1.1$
case, where $\alpha$ is the gravity-to-pressure ratio), and the 
other highly over dense ($\alpha=4.0$).
As intuitively expected, the core with the small over-density factor
loses equilibrium and contracts slowly (timescale
1 Myr), while the highly over dense core contracts in one tenth of
the time. Due to these very different time scales,  
the two cores develop significantly different chemical
compositions by the time their central densities reach values like those
of L1498 and L1517B. The $\alpha=1.1$ core has lost
most molecular species at the center, while the $\alpha=4.0$
core still retains a significant fraction of depletion-sensitive 
molecules in the gas phase. From this dichotomy, AHRC05 argue that the
$\alpha=1.1$
model approximates the evolution of quiescent, heavily depleted
cores like L1498 and L1517B, while the $\alpha=4.0$ model 
simulates dense, but not depleted cores like L1521E.

Our molecular survey of L1498 and L1517B can be used to test the 
predictions of the different chemical models of core contraction. 
We choose the AHRC05 model because it is the most complete
model available and because it presents predictions of both
the core velocity field and its chemical composition.
To test this model, we have run a series 
of Monte Carlo radiative transfer calculations
using the observed density profiles of L1498
and L1517B together with 
the abundance and velocity profiles predicted by the
AHRC05 models
at the time when the core central density 
reaches $n_0 = 1.5 \times 10^5$ cm$^{-3}$ (note that
AHRC05 quote densities of H nuclei while we use densities of H$_2$
molecules). At this time, the density profiles of the AHRC05 
models are close to those of L1498 and L1517B, although 
there are still factor-of-2 differences between the density profiles 
that should be kept in mind when comparing with our observations.

Concerning the velocity field, we find that the  $\alpha=4.0$
model predicts for NH$_3$ lines much broader than observed (0.38 km s$^{-1}$ 
versus 0.2 km s$^{-1}$, paper 1), and for species with central depletion
like C$^{18}$O and CH$_3$OH, the model predicts spectra with two peaks.
These peaks correspond to the front 
and the back sides of the core, and are separated by 0.4 km s$^{-1}$ because
each side is moving toward the center with a velocity of about 0.2 km s$^{-1}$
(Fig. 1e in AHRC05). Such broad, double-peaked profiles are not observed in
L1498 or L1517B, and they rule out a fast contraction model for
these cores. The $\alpha=1.1$ model, on the other hand, predicts narrower 
spectra, similar to those observed. In the absence of turbulence, this model
predicts the correct NH$_3$ linewidth toward the center 
for both L1498 and L1517B (0.2 km s$^{-1}$), but for species with central 
depletion, it predicts lines that have two peaks separated
by 0.15 km s$^{-1}$, and this is not observed.
Although these peaks can be blended into a single feature by adding a small
amount of turbulence, the lines are already broader than
observed and the extra turbulence further degrades the fit. For the
narrow-lined 
CH$_3$OH emission, for example, the AHRC05 model without turbulence
predicts lines that are broader than measured by 20\% in the case of
L1517B and by 70\% for L1498. This disagreement 
may be exaggerated by the imperfect match between the
density profiles of our cores and the AHRC05 models, but also
illustrates the fact that in the AHRC05 $\alpha=1.1$ model, densities 
like those of L1498 and L1517B are reached at relatively late 
stages, when the core contraction has started to accelerate. For
this reason, the model predicts that the L1498/L1517B phase
will only last 10\% of the core time life, or just $10^5$ yr, which is 
too short for a typical starless core (e.g., \citealt{lee99}).
The relatively fast contraction of the model is therefore not supported
by the observations. Adding a magnetic field can help to slow down the 
contraction, but this may introduce additional problems with the chemistry. 
A long period of slow contraction, for example 1 Myr,
would imply CO depletion at densities as low as $10^4$ cm$^{-3}$
\citep{leg83}, which is lower than observed. Accretion of new
(undepleted) material from the surrounding molecular cloud may 
help mitigate this difficulty. An additional problem with 
chemical models of magnetized clouds is that they predict
N$_2$H$^+$ to trace the quiescent and thermally supported core nucleus 
in contracting starless 
clouds such as L1544 \citep{she03}, in disagreement with 
observations \citep{cas02b}.

\begin{figure}[t]
\centering
\resizebox{\hsize}{!}{\includegraphics{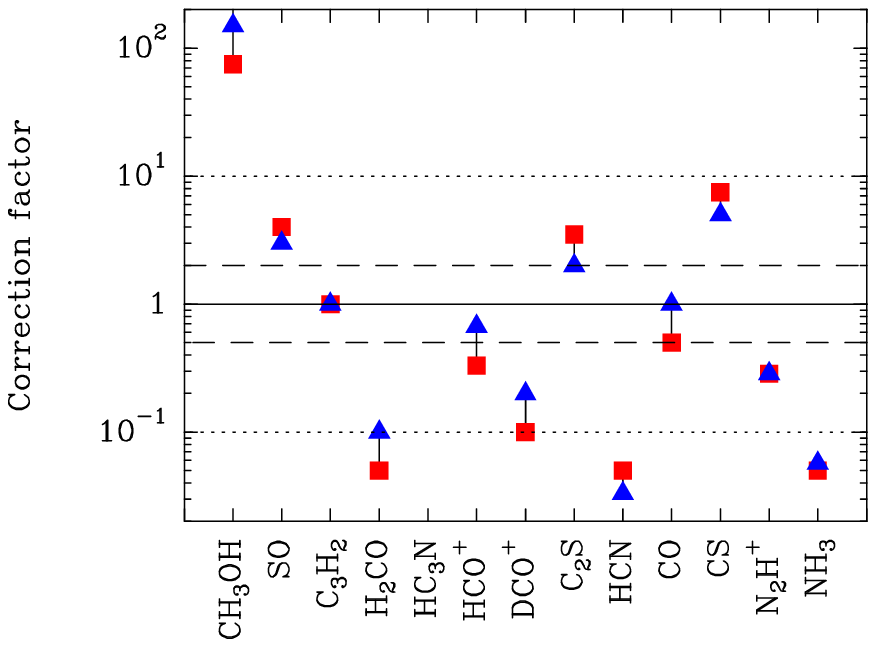}}
\caption{Correction factors needed to make the abundances
predicted by 
the \citet{aik05} $\alpha=1.1$ model reproduce the
observed intensities in the L1498/L1517B molecular
survey. The blue triangles
(L1517B data) and the red squares (L1498 data) 
indicate that order of magnitude
corrections are needed in a number of species. 
The comparison between model and
observations has been done at the fiducial outer radius
defined in section 4.2.
The dashed and dotted 
lines indicate factor of 2 and 10 corrections, respectively.
\label{fig7}}
\end{figure}

To test the AHRC05 abundance predictions, we now fix the velocity field of
each core to the best fit value determined in paper 1, and we use the Monte
Carlo code to estimate predicted radial profiles of integrated intensity for 
all species observed in L1498 and L1517B and 
predicted by model $\alpha=1.1$ ($n_0 = 1.5 \times 10^5$ cm$^{-3}$).
If the AHRC05 abundance profile does not fit
the observations at the fiducial outer radius defined in section
4.2 ($75''$ for L1498 and $55''$ for L1517B),
we estimate the global factor by which 
the model abundance needs to be multiplied to match the data
at that radius. This factor 
measures the deviation of model from the observations, 
and is shown in Fig. 7 for both the L1498 and L1517B models.
As expected from the good agreement
between the L1498 and L1517B abundances (section 5.1), the
correction factors derived from the two cores agree within a 
factor of two. This shows that the corrections are almost
independent of the exact structure of the core.

The correction factors in Fig. 7 cover more than three orders of
magnitude, and this indicates that some species are seriously
over or under predicted by the model. Considering
all the uncertainties of our fitting procedure, we take any 
correction smaller than a factor of two as a reasonable
match between model and data. This occurs for C$_3$H$_2$ (although
the model misses the observed outer abundance drop), HCO$^+$
(although L1498 requires a factor of 3 correction), and CO. 
Correction factors between 2 and 
one order of magnitude are considered ``clear deviations''
between model and data, and include SO, DCO$^+$, C$_2$S, CS,
and N$_2$H$^+$. Interestingly enough, three of these species
are S-bearing, and all of them are under predicted 
by a factor of a few (5-7.5 for CS, and 2-4 for SO and C$_2$S).
This suggests that the model overestimates the 
depletion of S in its initial atomic conditions by a similar factor.
Finally, correction factors larger than one order of magnitude 
indicate a ``serious deviation'' between model and data, and occur for
CH$_3$OH, H$_2$CO, HCN, and NH$_3$. Given its large value, 
the deviation of
H$_2$CO is likely real, although thin isotopologue observations are
needed to better quantify the error. On the other hand, 
the large under production of CH$_3$OH is not surprising given the 
lack of a known
gas-phase production mechanism for this molecule \citep{luc02}.
As for HCN, it is possible that the addition of photochemistry
to the model may help correct the overproduction of this molecule.

The need for order-of-magnitude corrections does not necessarily 
imply that the AHRC05 model fails to reproduce 
all aspects of the chemical structure in cores like L1498 and L1517B. 
Most abundance profiles, when scaled appropriately, reproduce the 
shape of the 
radial distribution of intensity, and this suggests that the model 
captures at least part of the process of molecular depletion
as the core contracts. To study how well
the AHRC05 model predicts the radial variation in the abundance of the
different species, we compare the {\em shape} of the observed intensity
profiles with the shape predicted by the model after correcting 
the abundances by the appropriate global factor. We do this by 
defining a ``concentration factor'' as 
the ratio between the intensity at the core center and the intensity
at a fiducial outer radius ($75''$ for L1498 and $55''$ for L1517B,
see section 4.2). Such a factor measures how centrally
peaked the emission is, and can be easily estimated both in the
model predictions and in the data. The ratio between the model and
data concentration
factors will equal one in a perfect match, will be larger than one
if the model under-predicts the central abundance drop (so its emission
is more centrally peaked), and will be less than 1 the other way around. 

\begin{figure}[t]
\centering
\resizebox{\hsize}{!}{\includegraphics{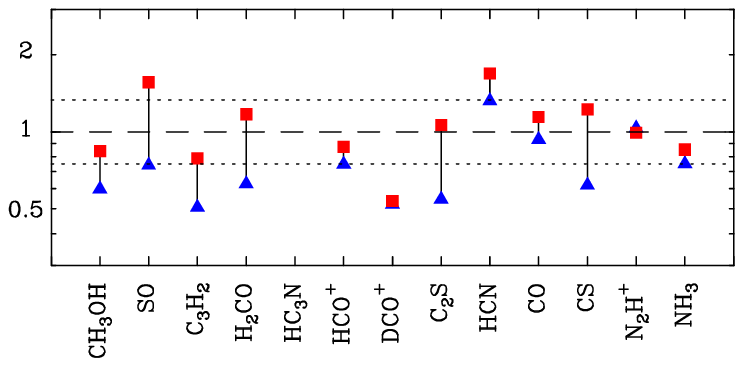}}
\caption{Comparison between the concentration ratios predicted
by the \citet{aik05} model and the observations of L1498 
(red squares) and L1517B (blue triangles).
A value equal to 1 indicates that the model predicts emission
with the same central
concentration as observed, a value larger than 1 indicates that
the model is more concentrated than the data (small central hole), 
and a value less than 1 indicates a model prediction 
flatter than the data (too large a hole). There is an overall
(factor of 2) agreement between model and data, although 
the L1517B data seems systematically lower than 1. The model
therefore overestimates the central hole seen in L1517B.
\label{fig8}}
\end{figure}

Figure 8 presents the ratio of concentration factors (model over data) in both
L1498 and L1517B for all available species. For
L1498, the AHRC05 model predicts on average the correct size of the 
central hole, and the average model-to-data ratio is approximately 1.
For L1517B, the model over predicts the size of the central
hole, as would be expected due to its smaller dimension
(section 5.1), and the mean ratio of concentration factors 
is about 0.75. This systematic over prediction for L1517B reflects 
a global problem modeling the core, so in the following discussion
we concentrate on the L1498 results.
As the figure shows, SO 
and HCN present the largest ratio of all, indicative that the model 
under predicts their central abundance hole by the largest factor.
This occurs in SO because this species has
a relatively large hole, while the AHRC05 model
predicts an average value. For HCN, the data show a relatively small
central hole, but the AHRC05 model predicts negligible depletion at
the time when the core has the central density of L1498 and L1517B. (Note
that the model correctly predicts a well defined, relatively smaller HCN hole
at later times.) DCO$^+$, on the other hand, is the species with smallest ratio
in the figure, indicating that the AHRC05 model predicts
a hole larger than observed. This is most likely due to an under prediction
of the central deuterium enhancement, as the model predicts the
correct depletion hole for the
main isotopologue HCO$^+$. Finally, 
the radial behavior of the two species without central freeze out, 
N$_2$H$^+$ and NH$_3$, is well predicted by the model after
dividing the N$_2$H$^+$ abundance by a factor of 3.5 and the NH$_3$
abundance by a factor of 20. This general over prediction
of N-bearing species probably results from an underestimate of the
binding energy to grains (see below), while the improved behavior of 
NH$_3$ arises from the new treatment of the N$_2$H$^+$ dissociative 
recombination thanks to the work by \citet{gep04}.

A final problem affecting most current chemical models (including AHRC05, 
but also \citealt{ber97}) is that they owe most of their success in
explaining the different depletion behavior of C-bearing and N-bearing species
to the assumption that CO and N$_2$ have significantly different binding 
energies to grains. Recent laboratory measurements by 
\citet{obe05} and \citet{bis06}, however, show that the two binding energies 
differ by less than 10\%, and this may seem insufficient to account
for the different depletion behavior of CO and N$_2$H$^+$
(note however that \citealt{aik01} find differential depletion
even when using similar binding energies for CO and N$_2$).
\citet{flo05}
have proposed an alternative explanation in terms of a lower sticking 
coefficient for N$_2$ or N. While the former alternative seems ruled 
out by laboratory measurements \citep{bis06}, the latter is
still a viable solution, although no laboratory measurements
exist yet to confirm or refute the idea. This remaining uncertainty 
in our understanding of the process behind the
differentiation of C-bearing and N-bearing species
--the most visible feature of core chemistry-- illustrates
how it is still premature to use current chemical models to assign
contraction ages to cores. A new generation of models is still
needed to fulfill the promise of providing core studies with 
a reliable chemical clock.

\subsection{How to best identify depletion (or its absence)}

\begin{figure}[t]
\centering
\resizebox{\hsize}{!}{\includegraphics{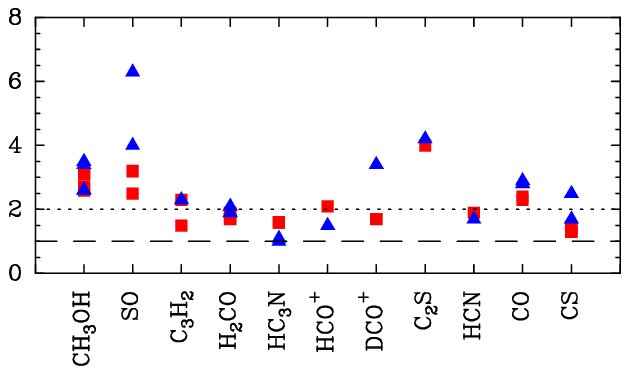}}
\caption{Effect of depletion on the different lines and transitions 
observed in the survey. The y-axis represents the ratio between the 
central intensities of a constant abundance model and the best 
fit model for each of the lines observed in L1498 (red squares)
and L1517B (blue triangles). Large values indicate a
strong effect of the central depletion in the emission
from the core center.
\label{fig9}}
\end{figure}

The simplest use of depletion as a qualitative indicator of 
core age is the classification of cores as chemically evolved if they 
show evidence for molecular depletion and as chemically young if they do not.
Most dense cores, in fact, suffer from severe depletion
(e.g., \citealt{bac02,taf02}), while only a minority
seem unaffected by it \citep{hir02,taf04b}. This suggests that 
young cores are rare, either because they are absent
from clouds or because an observational bias
limits our ability to recognize them in surveys.
In either case, it is of interest identifying the
most sensitive tracer of molecular depletion
to use it for systematic searches of young cores.
Choosing such a tracer 
requires some consideration. Sensitivity to depletion
depends not only on the relative size of the central
hole, but on the response of the emission
to dense gas and on possible optical depth effects.
In this section, we
use our radiative transfer analysis of L1498 and
L1517B to assess the effect of depletion on the emission
of different molecules under realistic core conditions, 
in order to find the species most sensitive to central depletion.

Strictly speaking, the sensitivity to depletion depends
not on the molecule but on its transitions. These may
greatly differ in critical density and optical depth,
especially in molecules with complex level structure. However, 
as we will see below, the results tend to agree within a 
factor of two. To measure quantitatively the sensitivity to
depletion of a molecular transition,
we compare the intensity toward the core
center predicted by our best-fit Monte Carlo model with 
the prediction from a model of constant abundance 
equal to the outer value in the best fit 
(solid and dashed lines in Figs. 3 and 4). The ratio
of the constant abundance result
over the depletion result measures
how much brighter a core with constant abundance 
would appear compared with the core suffering real depletion.

Figure 9 presents the depletion sensitivity ratio for all 
transitions observed toward L1498 (squares) and L1517B (triangles). 
For most molecules, the L1498 and L1517B values agree within a 
factor of two even when multiple transitions 
have been observed. As the figure shows, a sensitivity
factor of 2 (dashed line) is common for most species, indicating that
molecular depletion typically halves the expected
intensity towards the core center. Several species, however,
consistently present sensitivity ratios larger than 2. The 
SO lines reach the highest values of all, and have a mean factor 
of 4. This behavior is somewhat expected from
the systematically larger depletion radius of this molecule
(Section 5.2) and its moderately large dipole moment
(1.55 D), which makes it sensitive to the presence or
absence of SO in the inner core. Not surprisingly, the
highest SO ratios correspond to the JN=43--32 transition 
(138 GHz), which has an Einstein A coefficient larger than 
the other observed SO line, JN=32--21 at 99.3 GHz. Second in 
sensitivity to depletion is C$_2$S, with small scatter and also 
a mean value 
of 4. This molecule again combines a larger-than-average
depletion radius with lines of relatively large Einstein A.
In third position, and 
with a sensitivity ratio of about 3, lies CH$_3$OH, which also
presents a depletion radius slightly larger than average, 
sizable A coefficients, and intensities similar to those of the
SO lines. Interestingly, the molecule with largest depletion
radius, C$^{18}$O, has a sensitivity factor marginally better than 
2, which results from a combination of a low dipole
moment and moderate optical depth. The thinner C$^{17}$O (not
shown) is significantly more sensitive, although still 
suffers from contamination by the extended cloud.
SO, C$_2$S, and CH$_3$OH seem therefore the top three choices
for any systematic search of young, undepleted
cores.

\subsection{Origin of the line emission. Contribution functions}

Our L1498 and L1517B models reproduce the observed molecular emission,
so they are expected to provide good approximations to the core internal
excitation and radiation transfer. We can therefore use them
to investigate how the line emission of different molecules 
is produced inside the 
core, and how the emission propagates and is distorted by optical
depth effects as it travels towards the outside. Understanding
these processes is critical when interpreting observations of
similar systems for which a full radiative transfer analysis is
not available, and can only be done by examining the internal
properties of well-calibrated models.

\begin{figure*}[t]
\centering
\resizebox{15cm}{!}{\includegraphics{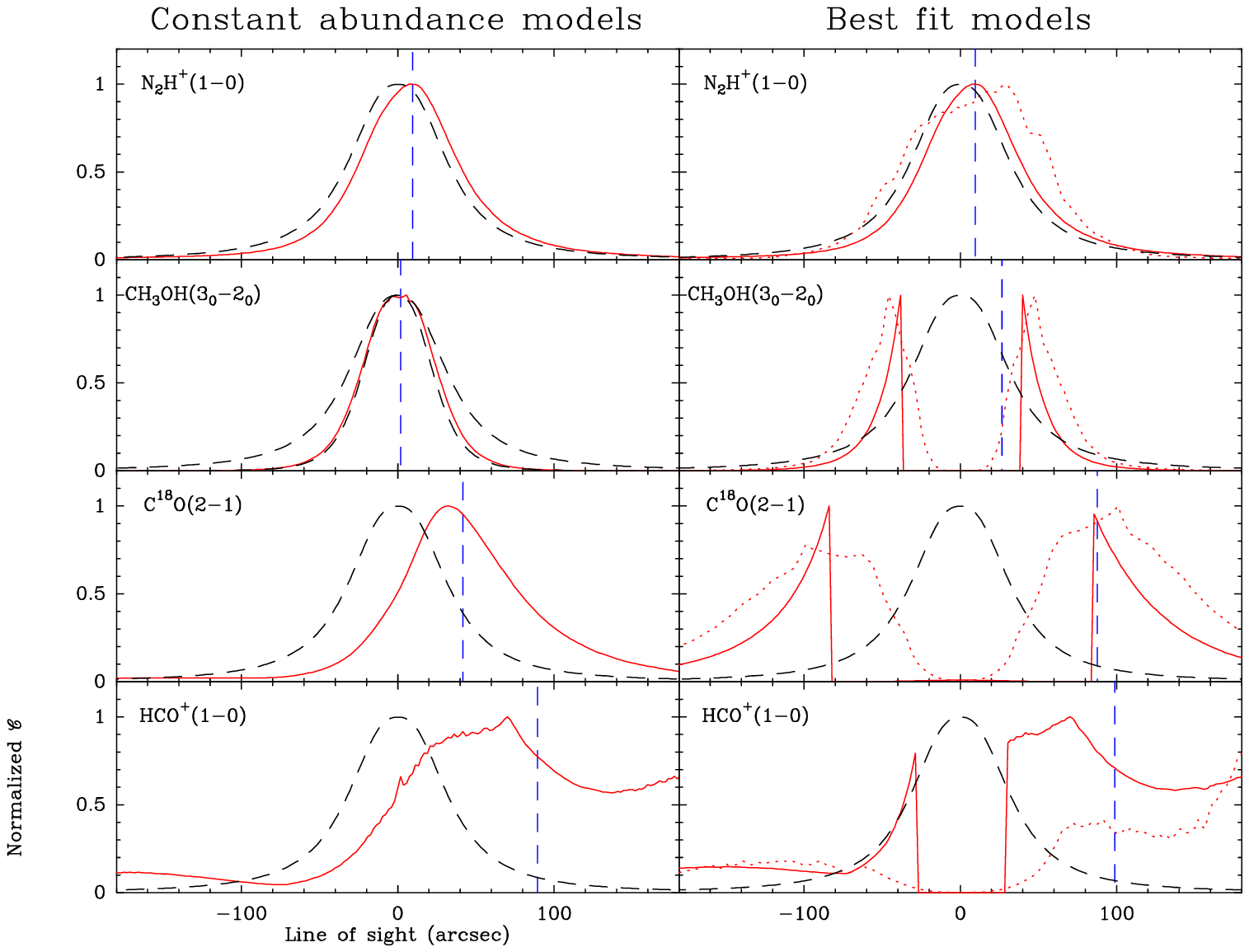}}
\caption{Normalized contribution functions of
representative species in L1517B (the observer is at large radius, i.e.,
to the far right of each panel). The red solid and dotted lines
are the contribution functions, while the black dashed lines indicate the 
normalized density profile shown for reference (the normalized square
of the density profile is also shown for CH$_3$OH). In the best fit
models (right column), the solid line corresponds to our simple
abundance
step function, while the dotted lines are the contribution functions
expected for the abundance profiles predicted by the \citet{aik05} models
(scaled appropriately to fit the L1517B data).
In each panel, the vertical dashed lines indicates the radius at which
half of the total emission is reached. Note how in the moderately thick 
C$^{18}$O(2--1) and extremely thick HCO$^+$(1--0) lines most of
the emission arises from the front outer layer of the core.
\label{fig10}}
\end{figure*}

The general problem of line formation in a core is complex 
because of the non linear nature of the radiative
transfer equation. In this section, we concentrate on
the question of how the different layers of a core contribute
to the emerging intensity, and on how different molecules can 
sample (or miss) the internal structure of the core. To 
quantify the discussion, we make use of the ``contribution function''
(CF) commonly used in the study of stellar atmospheres
(e.g., \citealt{gra05,mag86}). This function is derived from
the formal solution of the radiative transfer equation. According to
this solution, the emerging intensity from the core at a given
angle and frequency (without the background contribution) is
$$I_\nu^{\mathrm core}=\int_0^{\tau_m} S(\tau) \; e^{-\tau} d\tau =
\int_0^{l_m} j(r) \; e^{-\tau} dl,$$
where $S$ is the source function, $\tau$ the optical depth measured
from the core surface inwards, $\tau_m$ is the total core optical depth,
$j$ is the emissivity,
$dl$ is the line-of-sight element, and $l_m$ is the full line-of-sight 
length of the core. In this notation, the contribution function in
space units is
$${\mathscr C}(l) = j(r) \; e^{-\tau(l)}$$
(note that stellar atmospheres texts commonly define the CF as a
function of optical depth).

The emergent core intensity is the line-of-sight integral of ${\mathscr C}$, 
so the CF is true to its name in the sense that it measures the contribution
of a given core element to the observed line intensity. Strictly speaking, the CF 
depends on wavelength because different regions of the core may contribute
to different parts of the line profile. However, because of the small velocity
gradients in L1498 and L1517B (paper 1), and for the sake of simplicity,
here we will only deal with the frequency-integrated CF, which measures the 
contribution of a line-of-sight element to the integrated line profile. 
Also for simplicity, we will only study the central line of sight of the core,
although the generalization of the CF for non-zero impact parameters is 
straightforward.
With these assumptions, it can be easily shown that the CF of an optically thin
line (where ${\mathscr C} = j$) in an {\em isothermal} core
has two simple limits. If LTE applies (``high density''), 
the CF is proportional to the density $n(r)$, so the emergent intensity is 
proportional to the column density. If A$_{ul} \gg$ C$_{ul}$ (where 
A$_{ul}$ is the Einstein A coefficient and C$_{ul}$ is the collision
coefficient, ``low density'' case), the CF is proportional to n(r)$^2$, 
as each emerging photon is the result of a collision (which has a 
probability proportional to $n(r)^2$). In this case, the emergent 
intensity is proportional to the neutral analog of the
``emission measure.'' Monte Carlo tests using different species and transitions
under realistic core conditions show that
the CF for a thin line usually lies somewhere between the above two limits.

To illustrate the variety of CFs found during our radiative transfer 
modeling, we
present in Figure 10 a series of normalized CFs for different species in 
L1517B (plots for L1498 are similar). Each panel represents a cut 
along the central impact parameter of the core, has the density
peak at $r=0$, and assumes that the observer is located to the far right
of the plot (at $r \to \infty$); a normalized density profile in dashed lines
indicates the LTE optically thin limit of the CF. Constant abundance
models are presented in the left column and best-fit abundance models
appear on the right; we first discuss the constant abundance case.

As Figure 10 shows, 
the constant abundance model for N$_2$H$^+$(1--0) (and that of NH$_3$, 
not presented) has a CF that closely follows the density profile, with a
a slight shift to positive radius due to optical depth effects
(see below). This good behavior of the CF shows that
the  N$_2$H$^+$(1--0) emission responds linearly to density 
and therefore traces faithfully the core structure. To quantify this 
property, we compare the fraction of integrated CF that arises from 
the ``inner core'' (as defined within the half maximum density radius) 
to the fraction of gas column density contained in the same region (0.62 
for L1517B). For both the N$_2$H$^+$ and NH$_3$ constant abundance models, 
we find that the inner core contributes to the emission a percentage 
indistinguishable from the ideal 62\% value. In contrast, the CF of 
CH$_3$OH(3$_0$--2$_0$) in a constant abundance model (Fig. 10,
second left panel) 
follows closely the $n(r)^2$ curve, and is therefore significantly 
biassed toward the high density gas. For this line, the inner core 
contributes 85\% to the emergent emission, which is
almost 40\% more than the column density fraction.

The CF for the C$^{18}$O constant abundance model shows signs of 
non negligible optical depth in the form of a forward shift
of the peak. This shift indicates that most of the observed 
C$^{18}$O(2--1) emission arises from the front part of the core, and 
that the emission from the back is heavily attenuated as it travels 
through the core. In fact, only 15\% of the emerging radiation 
originates from the backside ($r<0$), and the inner core contributes
with only 38\% of the emission (a 40\% decrease from the ideal value).
C$^{18}$O, therefore, is a biassed tracer of the core emission despite 
its LTE excitation, although most of these problems can be minimized 
using C$^{17}$O transitions (but see below for models with depletion). 
As a final example of a constant abundance model, 
we consider the heavily self absorbed HCO$^+$(1--0) line, although
similar results are obtained for CS(2--1). The emission from this line
is so dominated by the outer core that half of it originates from radii
larger than $90''$, and the inner core only contributes 17\% of the emergent
intensity (23\% in the case of CS(2--1)). These numbers illustrate the 
difficulty of modelling
self absorbed lines like HCO$^+$(1--0), CS(2--1), or HCN(1--0):
the region that dominates the line emission is poorly constrained by the
continuum observations that are the basis of our density profile,
and its structure is simply guessed using an extrapolation of the central
core parameters. These very thick lines are therefore excellent tracers
of the outer velocity fields (like infall), but they become
poor tools for deriving any line of sight structure, even of the infall
velocity, unless there is a very
accurate description of the outer core and the surrounding cloud.

The presence of depletion introduces a
substantial distortion in the CF. This is illustrated by the right-hand
panels of Figure 10, which are based on the best-fit abundance profiles
of section 4.
The N$_2$H$^+$ CF is of course the same as in the constant abundance
case because no abundance gradient was found for this molecule. The 
CH$_3$OH, C$^{18}$O, and HCO$^+$ CFs, on the other hand, present sharp
drops near the core center that correspond to 
the abundance holes found for these species. Because of these 
drops, the CF in the inner core vanishes or becomes so small that a
negligible part of the emergent emission arises from the dense gas:
5\% in HCO$^+$(1--0) and less than 1\% in both CH$_3$OH(3$_0$--2$_0$) 
and C$^{18}$O(2--1). The spectra towards the core center for these
species is therefore overwhelmed by gas outside the core, mostly from
an innermost ring if the emission is not very thick (the case of CH$_3$OH
and C$^{18}$O) or from the front part of the core in the thicker HCO$^+$(1--0)
line. Although it can be argued that the abundance drops in our models
are artificially steep because of the use of a step function,
theoretical considerations of core chemistry still predict
very sharp abundance decreases. To test the effect of smoother, but 
realistic abundance gradients, we have produced a set of CFs using the
$\alpha=1.1$
model of \citet{aik05} corrected by the
appropriate scaling factors to match the observed
radial profile. The new CFs (shown as dotted 
lines in the right hand panels of Fig. 10) again present very steep holes toward
the core centers and predict inner-core contributions to the emerging 
intensity of 15\% (CH$_3$OH) and less (C$^{18}$O and HCO$^+$). The inner
core contribution to the emission for these and other depleted
species is therefore a minor fraction of the total emergent intensity.

The above small sample of CFs illustrates some of the basic phenomenology 
of line formation in cores like L1517B. As the plots show, even in the 
simple case of constant abundance, the emission is not generated at
a ``$\tau=1$ surface,'' but arises from a broad range of radii that
in the simplest case (thin LTE) has a weight that 
mimics the core density distribution, and that
in a thick line is strongly biassed toward the front part of the core.
When realistic abundance profiles are considered, 
if the species is sensitive to depletion, 
all hope of tracing the inner core is lost no matter how much 
spatial resolution is achieved. Only depletion-resistant species like 
N$_2$H$^+$ and NH$_3$ trace the inner gas, at least to densities of
a few $10^5$ cm$^{-3}$. Among these tracers, N$_2$H$^+$(1--0)
seems specially reliable because of a combination of constant abundance,
reasonably low optical depth (helped by the hyperfine structure), and
well-behaved excitation. Ignoring the NLTE ratios between the J=1--0
hyperfine components (a 10\% effect most likely due to a population 
redistribution), this line consistently produces a CF very close in
shape to the density profile. This is somewhat surprising given the
significant radial gradient in the excitation temperature 
of this line (more than a factor of 2 change between center and edge
in L1517B), and it results from a population of the J=1 level that follows
closely the density law (the population of higher N$_2$H$^+$ levels
falls faster than the density law, and the CFs of higher transitions
is closer to $n(r)^2$). This fortunate behavior of the N$_2$H$^+$ 
molecule explains the striking similarity between the mm continuum and 
N$_2$H$^+$(1--0) maps in many starless cores (e.g, 
\citealt{cas99,taf02,bac02,cra05}, this work),
and makes the 1--0 line the ideal choice 
for studying core interiors as long as its optical depth remains low.

\section{Conclusions}

We have presented observations of 13 molecular species (plus
a number of isotopologues) toward the Taurus-Auriga starless
cores L1498 and L1517B. Combining Monte Carlo radiative transfer
modeling with a physical description of the cores, 
we have derived a self-consistent set of molecular
abundance profiles. From the analysis of these profiles,
we have reached the following main conclusions.

1. Most abundance profiles can be described with simple step functions
having a constant value in the outer layers and a central hole. 
In both L1498 and L1517B, we find central abundance drops for
CO, CS, CH$_3$OH, SO, C$_3$H$_2$, HC$_3$N, C$_2$S, HCN, H$_2$CO,
HCO$^+$, and DCO$^+$, while N$_2$H$^+$ and NH$_3$ are found to remain 
in the gas phase even at the highest densities. The central 
abundance drops suggest that most species
freeze out onto the cold dust grains at the centers of the two cores.

2. For most species, the abundances estimated in the outer layers of
L1498 and L1517B agree with each other within a factor of 2, while the
central abundance holes are systematically larger in L1498 by a factor 
of about 1.5. This suggests that the two cores
have contracted from gas of similar composition, but that they
are following different contraction histories or that they are
at different evolutionary stages.

3. Not all species with depletion present central holes of the same
size. Molecules like DCO$^+$, HCN, and HC$_3$N systematically have
smaller than average central holes, indicating
that they survive in the gas phase up to slightly higher densities
than the rest of the depleting molecules. Species like SO, C$_2$S,
CH$_3$OH, and 
probably CO, on the other hand, present larger than average abundance
holes, and seem therefore to deplete at lower densities.

4. L1498 and L1517B seem to have chemical compositions similar to 
other Taurus cores like TMC1, although no other multi-molecular
survey of core chemistry that includes depletion
is available for a detailed comparison. From a review of
different abundance determinations in the literature, it seems
that the chemical composition of L1498 and L1517B is rather
typical of the Taurus starless core population. The abundances
determined in this work can therefore be used as a 
reference for future modeling.

5. A comparison with the recent (and most up to date) model
of core chemistry by \citet{aik05} shows a combination of
successes and failures. For most molecules, the model predictions 
need to be corrected by factors of 5 or even 10 in order to
match the observed intensities. The model, on the other hand,
predicts reasonably well (factor of 2) the size of the central 
depletion hole for most molecules. However, it fails
predicting the observed variety of hole sizes among molecules.

6. By comparing observed central intensities and 
intensities predicted by constant abundance models, we
have studied the relative sensitivity of the different molecules
and transitions to the presence of central depletion. We find that
SO and C$_2$S are the most sensitive indicators of molecular
depletion, and therefore constitute the best choices when searching 
for very young cores.

7. We have used radiative transfer modeling and realistic abundance
profiles to study how molecular lines originate at the core interior.
We find that the ``contribution function'' provides an excellent tool
to visualize and quantify the formation of the line profile from
the different parts of a core. Using this function, we show that
the N$_2$H$^+$(1--0) line is the most faithful tracer of the internal
core structure as long as it remains optically thin. Species
sensitive to freeze out only trace the edges of the central depletion
hole (if thin) or the outer front part of the core (if thick).

\acknowledgements
We thank the staffs of the IRAM 30m and FCRAO 
telescopes for support during the observations, 
Claudia Comito for providing us with the HC$_3$N(4--3)
data, Yuri Aikawa for useful comments on chemical modelling,
Kaisa Young and Jeong-Eun Lee
for information on their H$_2$CO data of L1498, and
the referee, S\'ebastien Maret, for a thorough and useful review.
MT acknowledges partial support from grant AYA2003-7584, and
PC and CMW acknowledge support from the MIUR grant ``Dust particles 
as factor of galactic evolution.''
This research has made use of NASA's Astrophysics Data System
Bibliographic Services and the SIMBAD database, operated at CDS,
Strasbourg, France.

\appendix

\section{Molecular parameters and convergence tests}

{\bf C$_2$S.}
Dicarbon monosulfide is a linear radical with two unpaired
electrons (electron spin S=1), and therefore fine structure
in its spectrum. Its rotational levels are characterized by 
N and J, the quantum numbers of rotational angular momentum and 
total angular momentum, respectively (e.g., \citealt{wol97}).
We model its radiative transfer using the
C$_2$S molecular parameters from 
the JPL catalog \citep{pic98} together with the C$_2$S-H$_2$
collision rates from \citet{wol97}
(downward rates for 10K together with upward rates
derived from detailed balance). We include
all energy states up to approximately 60 K, which results 
in a total of 37 levels and 79 transitions.

{\bf CH$_3$OH.}
Methanol is a slightly asymmetric top with a hindered internal 
rotation \citep{tow55}.
It consists of two symmetry species denoted A and E (e.g., 
\citealt{lee73}) whose rotation energy levels are characterized by 
their total angular momentum J and its component along the
near symmetry axis k (an additional $+$ or $-$ is required 
for the A species if k$\ne0$). For our calculations, we have
used the energy and line strength determinations in 
the JPL catalog \citep{pic98} together with the recent CH$_3$OH-para H$_2$ 
collision
rates of \citet{pot04} (downward rates for 10K together with upward rates
derived from detailed balance). For both species, we include
all energy states up to approximately 70 K, which results
in a total of 31 levels and 92 transitions for E-CH$_3$OH,
and 27 levels and 66 transitions for A-CH$_3$OH.

{\bf c-C$_3$H$_2$.}
Cyclopropenylidene, a ring molecule, is an oblate asymmetric top
with ortho and para species \citep{vrt87}. Its rotational
levels are characterized by the total angular momentum J and by
K$_{-1}$ and K$_{+1}$, its two projections on the symmetry axis
for the prolate and oblate symmetric top limits \citep{tow55}.
For our calculations, we have used the C$_3$H$_2$ 
parameters in the JPL catalog
\citep{pic98} together with the C$_3$H$_2$-He collision rates 
for 10 K from 
\citet{ave89} (multiplied by $2^{1/2}$ to simulate collisions with
H$_2$, as recommended by the authors). Only transitions of 
ortho C$_3$H$_2$ were observed, so our modeling is restricted to 
that species. This modeling includes all states with available 
collision rates (up to $E_u=41$ K): 16 levels and 32 transitions.

{\bf H$_2$CO.}
Formaldehyde is a slightly asymmetric prolate rotor with ortho and
para species, and its rotation levels are characterized (like those
of C$_3$H$_2$) by J, K$_{-1}$, and K$_{+1}$. For our calculations,
we have used the H$_2$CO parameters from the on-line
Cologne Database for Molecular Spectroscopy (CDMS, \citealt{mul01}),
together with the 10 K H$_2$CO-He collision rates of \citet{gre91}. 
These collision rates were multiplied by 2.2 to convert them 
into H$_2$CO-H$_2$ rates, as recommended by \citet{gre91}.
We only observed lines of ortho H$_2$CO,
so only this species has been modeled. For this, we have
taken into account all energy states with 
energy below 115 K, resulting in a calculation with 14 levels
and 19 transitions.

{\bf HC$_3$N.}
Cyanoacetylene is a linear molecule. Its rotational
levels have hyperfine (hf) structure due to the nuclear
spin of the nitrogen atom, so they are characterized
by the quantum numbers J and F. For our radiative
transfer calculation we have used the molecular
parameters from \citet{laf78} and the collision 
rates with He of \citet{gre78} (multiplied by 
$2^{1/2}$ to simulate collisions with
H$_2$, as recommended by the authors). The calculation 
treats the hf structure in a simplified manner 
by solving the equations of statistical equilibrium 
for the pure rotational levels after
dividing by 3 the effect of trapping, 
to simulate the decrease of optical depth
due to the three-fold hf splitting
(we find this to be a minor effect).
The resulting population of
each rotational level is then divided between the
hf sub levels assuming LTE to compute the emergent
spectrum (see \citealt{taf02}
for a discussion of a similar approach with N$_2$H$^+$). All
24 levels up to an energy of 130 K are considered
(23 transitions). 

{\bf HCO$^+$ and isotopologues.}
The formyl cation HCO$^+$ and its isotopologues H$^{13}$CO$^+$ and
DCO$^+$ are linear ions. We have modeled their radiative transfer
using the parameters in the JPL catalog together 
with the 10 K HCO$^+$-H$_2$ 
collision rates of \citet{flo99}. Our calculation 
considers a set of 7 levels (6 transitions), 
up to an equivalent energy of 120 K. To take into 
account the small additional broadening due to hyperfine structure 
in H$^{13}$CO$^+$, we have artificially 
broadened this line by an extra 0.133 km s$^{-1}$
(see \citealt{sch04}).

{\bf HCN and H$^{13}$CN.}
Hydrogen cyanide is a linear molecule with hyperfine
structure due to the nuclear spin of the nitrogen atom. 
We have modeled its radiative transfer using the molecular 
parameters in the CDMS together with the HCN-He collision
rates of \citet{mon86} (for J up to 4) and the HCN-He rates
of \citet{gre74} (remaining levels 
up to J$=7$). A factor of 2 was used to convert the He collision
rates into H$_2$ rates. The hyperfine structure of this species
(and that of its isotopologue) has been considered explicitly
in the calculations, although overlap effects were ignored
and the collision rates between hf sub levels with J$>4$
were estimated from the \citet{gre74} rates
assuming independence of the F number \citep{gui81}. A total
of 22 levels and 39 transitions were included, corresponding to 
all levels with energy lower than 119 K (up to J=7).

{\bf SO.}
Sulfur monoxide is a diatomic molecule 
with two unpaired electrons and therefore a fine structure
similar to that of C$_2$S. We have modeled its radiative transfer
using the molecular parameters in the CDMS together with the 
SO-He collision rates from \citet{liq05}. Following the
authors's recommendations, 
a factor of 1.38 was used to convert the SO-He collision
rates into SO-H$_2$ rates.
All 24 levels up to an energy of 85~K and their corresponding
72 transitions were included in the model.

{\bf Convergence tests.} 
When solving the radiative transfer,
we divide the core in
200 equally spaced shells. We normally use 2000 photon packages
to simulate the radiation, and allow 40 iterations for the solution
to converge (no radiation reference field was used, see \citealt{ber79}
for details on the code). To check the convergence of the
solution, several tests were carried out. Independence
on the initial level population was tested by comparing the results from
runs that started with Boltzmann equilibrium at 10 K (kinetic
temperature) and at 2.7 K (cosmic background); no differences in the
final solution were found. Independence on the internal Monte Carlo
parameters was tested by repeating the best fit solutions using
5000 photon packages and 300 iterations. Again, no appreciable
differences were found in the predicted intensities.

\section{An additional gas component toward the L1498 core}

\begin{figure}[t]
\centering
\resizebox{\hsize}{!}{\includegraphics{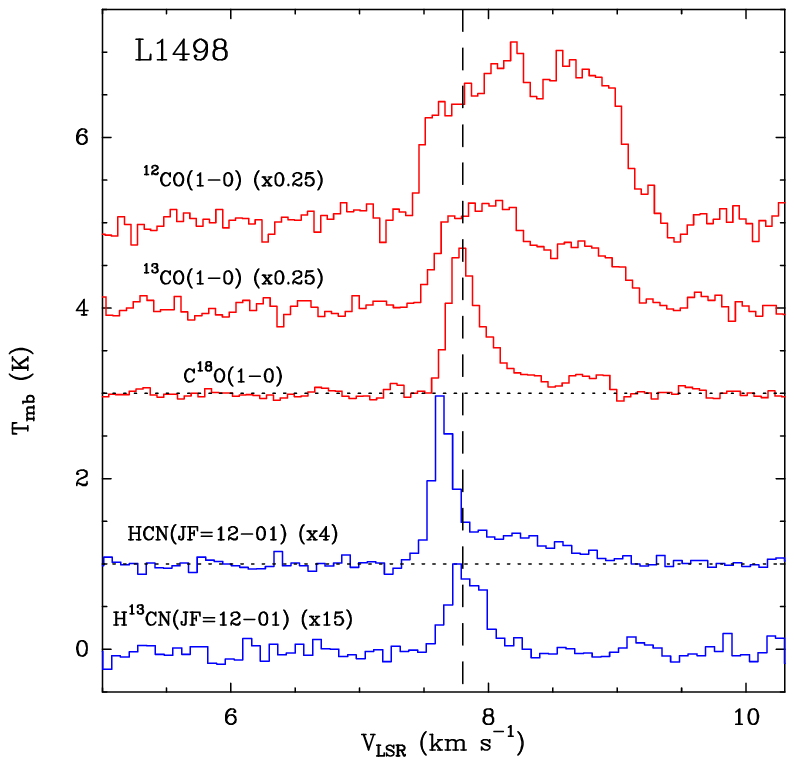}}
\caption{Spectra towards L1498 illustrating the presence of 
a red gas component. The bottom (H$^{13}$CN) line shows
that the LSR velocity of the cloud is about 7.8 km s$^{-1}$
(vertical line). The HCN line (next up) shows a deep red self-absorption
and a low level red wing, while the CO isotopologue lines 
show an increasing contribution of the red component with
abundance. The coincidence of the red component with the
absorption in HCN (also HCO$^+$ and CS) suggests that
the component lies in front of the L1498 core.
\label{fig11}}
\end{figure}

The line of sight toward the L1498 core is complex.
In addition to the narrow-line component that corresponds to the
core studied in this paper, spectra of abundant species like 
$^{12}$CO and $^{13}$CO show extra emission toward the red extending up
to about V$_{LSR}$ = 9.5 km s$^{-1}$, which is 1.7 km s$^{-1}$ higher than 
the core velocity (Fig. B.1). This additional component has a complex spatial
distribution (see also \citealt{kui96}). 
At low velocities ($\sim 8.2$ km s$^{-1}$), it lies toward
the south west of the core (as also found by \citealt{lem95}, while at high 
velocities ($\sim 9$ km s$^{-1}$), it appears toward the NW. This
distribution around the L1498 core suggests some 
relation with the dense gas, although the velocity difference implies 
that the red gas is not bound to the core. From the C$^{18}$O(1--0) spectrum 
in Fig. B.1 and assuming an excitation temperature of 5-10 K together
with a ``standard'' C$^{18}$O abundance of 1.7 10$^{-7}$ \citet{fre82},
we deduce an approximate H$_2$ column density for this component
of 2 10$^{21}$ cm$^{-2}$ (less than 10\% of the core column density).

Although the red component does not affect the narrow optically thin
lines used in our abundance determinations, it can distort
the shape of very thick spectra like HCN(1--0), shown in
Fig. B.1. These spectra were also modelled in section 4º by assuming 
standard isotopic ratios, and they sometimes required anomalous abundances
in the outer core (section 4.2). As stressed before, these optically
thick lines were only fitted to prove self consistency, and not to
determine the conditions of the outer core layers. To model  
these lines, however, some assumptions had to be made about the red 
component, and a main one was its location along the line of sight with
respect to the L1498 core. A foreground position was preferred based on 
hints from several optically thick spectra, like the HCN(JF=12--01) line 
shown in Fig. B.1. This spectrum shows a deep self absorption that removes
completely the red part of the line and replaces it by a low-level plateau that
continues smoothly into the red-component range (from V$_{LSR}$ 
8 to 9 km s$^{-1}$). Such a feature is best understood if the red component
contributes to the absorption of the core emission and therefore arises 
from gas located in front of the core. To model such a component, we have 
red-shifted
the front outer core layers of L1498 (``envelope'' in paper 1) by 0.35 km
s$^{-1}$
and broadened them to $\Delta V = 0.625$ km s$^{-1}$
(note that because of a typo, Table 3 in paper 1 gives the wrong
envelope velocity, which should be 8.15 km s$^{-1}$); such a simple
parameterization reproduces approximately the shape of most optically 
thick lines. 

\section{Comparison with published data}

\begin{figure}[h!]
\centering
\resizebox{\hsize}{!}{\includegraphics{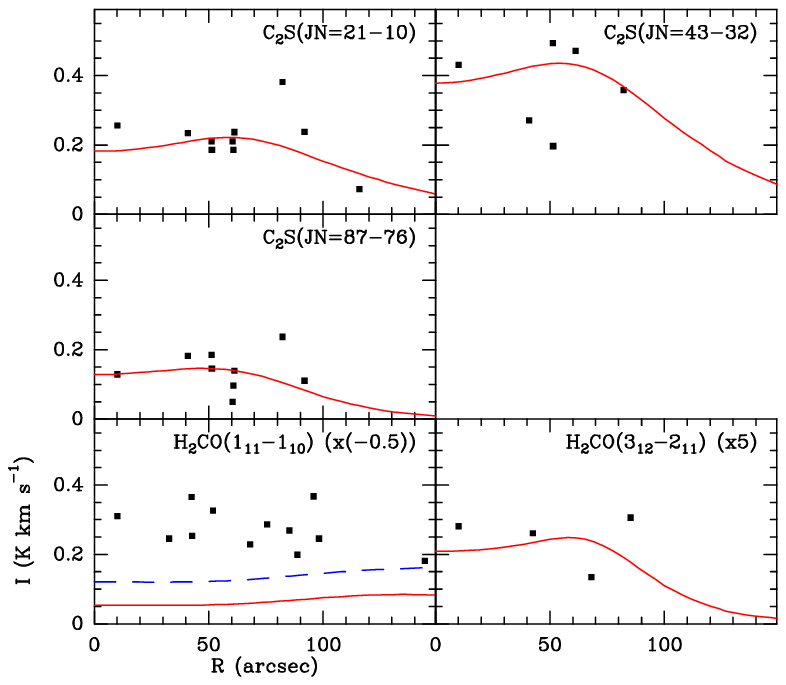}}
\caption{Comparison of L1498 data taken from the literature (solid squares)
with the predictions of the best fit model of section 4.2 (red lines). The top
three panels show that the model fits well the 
C$_2$S radial profiles generated from the data of \citet{wol97}. The
bottom two panels show that the model fits well the 
H$_2$CO($3_{12}$--$2_{11}$) line of \citet{you04}, but fails to
fit the H$_2$CO(1$_{11}$--1$_{10}$) line (from the same authors), which
appears in absorption against the cosmic background. To illustrate
the sensitivity of the 1$_{11}$--1$_{10}$ absorption to the core
low density gas, a model with a backside envelope is shown in
blue dashed lines (see text).
\label{fig12}}
\end{figure}

We have compared our L1498 model predictions with 
data for this core available in the literature (no
similar data were found for L1517B). \citet{wol97}
have presented observations of L1498 in three C$_2$S transitions,
JN=21--01 (22 GHz), JN=43--32 (45 GHz), and JN=87--76 (94 GHz), made
with the telescopes of NASA's Deep Space Network and the FCRAO, 
and having spatial resolutions of about $50''$. From the linewidths
and peak intensities reported by these authors in their
Table 2, we have reconstructed the line 
integrated intensities for all positions available, and with them, 
we have generated the radial profile of emission for each transition. 
When we compare these profiles with the prediction from the best abundance
model derived in section 4.2, we find the excellent match illustrated
by the three top panels in Fig. C.1. This match shows that our model
predictions are fully consistent with the observations of 
\citet{wol97}.

Other L1498 data of interest for our study are those
of \citet{you04}, who have observed L1498 in
H$_2$CO(1$_{11}$--1$_{10}$) (6 cm) with the Arecibo
telescope and in H$_2$CO($3_{12}$--$2_{11}$) (1.3 mm) with 
the CSO (spatial resolutions of $60''$ and $32''$).
To reconstruct the integrated intensities 
of these lines, we have used the values reported by the authors 
in their Table 2 (1.3 mm line) and data kindly provided by
Kaisa Young and Jeong-Eun Lee (6 cm line). With them, 
we have generated a set of radial profiles, and in 
Figure C.1 we compare them with the predictions from 
the best fit model of section 4.2. As the figure shows, the model
reproduces well
the $3_{12}$--$2_{11}$ emission, which is sensitive to the high density
gas inside the core. The model, however, fails to reproduce the
observed intensity of the 1$_{11}$--1$_{10}$ transition,
although it agrees with the data in producing an absorption line
against the cosmic background.
The poor fit occurs because the
absorption depends sensitively on the low density gas outside the core, and
this gas is not well parameterized by our extrapolation of the inner core
model (section 4.2). To illustrate this sensitivity to the low density gas,
we have added to the L1498 model an envelope in the back that has 
the same properties as the front envelope used to 
simulate the red shifted component (appendix B). This back envelope
doubles the 1$_{11}$--1$_{10}$ absorption, and brings the model
prediction closer to the observations (blue dashed lines in
Fig. C.1). Such strong sensitivity to the
low density gas shows that the central core is close to invisible in 
the 1$_{11}$--1$_{10}$ transition (note the flat radial profile), and that
improving the fit requires a re-parameterization of the envelope, not the 
core. Given the few observational constrains of the low density gas,
such a parameterization lies outside the scope of this work.

\bibliographystyle{apj}

\end{document}